\documentclass[11pt]{article}

\usepackage[x11names]{xcolor}

\usepackage{amsmath,amssymb,amsthm}

\usepackage[looser]{newpxtext}
\usepackage[tt=false,sf]{libertine}
\usepackage[varbb,smallerops]{newpxmath}
\usepackage{fullpage}
\usepackage{authblk}

\usepackage[colorlinks=true,citecolor=blue!65!black,linkcolor=red!65!black,backref=page]{hyperref}

\usepackage[nameinlink]{cleveref}

\newtheorem{theorem}{Theorem}[section]

\newtheorem{lemma}[theorem]{Lemma}

\newtheorem{observation}[theorem]{Observation}
\newtheorem{remark}[theorem]{Remark}

\newtheorem{problem}[theorem]{Problem}
\newtheorem{example}[theorem]{Example}

\usepackage{subfig}

\usepackage{cite}

\usepackage{comment}
\usepackage{multirow,multicol,dcolumn,booktabs,colortbl,hhline}
\usepackage{mathtools}
\usepackage{xspace}
\usepackage{bm}
\usepackage{framed}
\usepackage{fancybox,ascmac}
\usepackage{wrapfig}

\usepackage{tikz}
\usetikzlibrary{positioning}
\usetikzlibrary{decorations.markings}
\usetikzlibrary{shapes.geometric}
\usetikzlibrary{arrows.meta}
\usetikzlibrary{arrows,automata,decorations.pathmorphing,fit,positioning,arrows.meta,calc}
\usetikzlibrary{arrows,automata,decorations.pathmorphing,fit,positioning,arrows.meta,calc,decorations.text}
\usetikzlibrary{patterns}
\tikzstyle{mynode}  =[draw, circle, minimum size=0.9cm, inner sep=0pt, ultra thick, fill=white, font=\LARGE]
\tikzstyle{mymynode}=[draw, circle, minimum size=0.4cm, inner sep=0pt, ultra thick, fill=white]
\tikzstyle{mytoken} =[draw, circle, minimum size=0.9cm, inner sep=0pt, ultra thick, fill=black, text=white, font=\LARGE]
\tikzstyle{myactive}=[draw, circle, minimum size=0.9cm, inner sep=0pt, ultra thick, fill=white, pattern=crosshatch, pattern color=gray, font=\LARGE]
\tikzstyle{myedge}  =[ultra thick]
\tikzstyle{mylabel} =[font=\normalsize, inner sep=0]

\usepackage{nameref}
\crefname{theorem}{Theorem}{Theorems}
\crefname{lemma}{Lemma}{Lemmas}
\crefname{claim}{Claim}{Claims}
\crefname{remark}{Remark}{Remarks}
\crefname{observation}{Observation}{Observations}
\crefname{proposition}{Proposition}{Propositions}
\crefname{corollary}{Corollary}{Corollaries}
\crefname{example}{Example}{Examples}
\crefname{appendix}{Appendix}{Appendices}
\crefname{section}{Section}{Sections}
\crefname{equation}{Eq.}{Eqs.}
\crefname{algorithm}{Algorithm}{Algorithms}
\crefname{figure}{Figure}{Figures}
\crefname{table}{Table}{Tables}

\usepackage{algorithmicx}
\usepackage[ruled,section]{algorithm}
\usepackage[noend]{algpseudocode}
\algnewcommand\And{\; \textbf{and} \;}
\algnewcommand\Or{\; \textbf{or} \;}
\algnewcommand\To{\; \textbf{to} \;}
\algnewcommand\Continue{\textbf{continue}}
\algnewcommand\Not{\textbf{not}}

\algrenewcommand\textproc{\textsl}

\renewcommand{\geq}{\geqslant}
\renewcommand{\leq}{\leqslant}
\renewcommand{\mid}{:}

\newcommand{\bbN}{\mathbb{N}}

\newcommand{\calA}{\mathcal{A}}

\newcommand{\calF}{\mathcal{F}}

\newcommand{\calS}{\mathcal{S}}
\newcommand{\calT}{\mathcal{T}}

\newcommand{\cP}{\textup{\textsf{P}}\xspace}

\newcommand{\NP}{\textup{\textsf{NP}}\xspace}

\newcommand{\PSPACE}{\textup{\textsf{PSPACE}}\xspace}

\newcommand{\Wone}{\textup{\textsf{W}[1]}\xspace}

\newcommand{\TJ}{{\textbf{TJ}}\xspace}
\newcommand{\TAR}{{\textbf{TAR}}\xspace}
\newcommand{\TJN}{{\textbf{TJN}}\xspace}

\newcommand{\prb}[1]{\textup{\textsc{#1}}\xspace}

\DeclarePairedDelimiter\floor{\lfloor}{\rfloor}
\newcommand{\isep}{\mathrel{..}\nobreak}

\newcommand{\convexpath}[2]{
[   
    create hullnodes/.code={
        \global\edef\namelist{#1}
        \foreach [count=\counter] \nodename in \namelist {
            \global\edef\numberofnodes{\counter}
            \node at (\nodename) [draw=none,name=hullnode\counter] {};
        }
        \node at (hullnode\numberofnodes) [name=hullnode0,draw=none] {};
        \pgfmathtruncatemacro\lastnumber{\numberofnodes+1}
        \node at (hullnode1) [name=hullnode\lastnumber,draw=none] {};
    },
    create hullnodes
]
($(hullnode1)!#2!-90:(hullnode0)$)
\foreach [
    evaluate=\currentnode as \previousnode using \currentnode-1,
    evaluate=\currentnode as \nextnode using \currentnode+1
    ] \currentnode in {1,...,\numberofnodes} {
-- ($(hullnode\currentnode)!#2!-90:(hullnode\previousnode)$)
  let \p1 = ($(hullnode\currentnode)!#2!-90:(hullnode\previousnode) - (hullnode\currentnode)$),
    \n1 = {atan2(\y1,\x1)},
    \p2 = ($(hullnode\currentnode)!#2!90:(hullnode\nextnode) - (hullnode\currentnode)$),
    \n2 = {atan2(\y2,\x2)},
    \n{delta} = {-Mod(\n1-\n2,360)}
  in 
    {arc [start angle=\n1, delta angle=\n{delta}, radius=#2]}
}
-- cycle
}

\title{On Reconfigurability of Target Sets}
\author{Naoto Ohsaka\thanks{
CyberAgent, Inc., Tokyo, Japan.
\href{mailto:ohsaka\_naoto@cyberagent.co.jp}{\texttt{ohsaka\_naoto@cyberagent.co.jp}}; \href{mailto:naoto.ohsaka@gmail.com}{\texttt{naoto.ohsaka@gmail.com}}
}}
\date{\today}

\begin{document}

\maketitle


\begin{abstract}
We study the problem of deciding reconfigurability of target sets of a graph.
Given a graph $G$ with vertex thresholds $\tau$,
consider a dynamic process in which vertex $v$ becomes activated once at least $\tau(v)$ of its neighbors are activated.
A vertex set $S$ is called a \emph{target set} if all vertices of $G$ would be activated when initially activating vertices of $S$.
In the \prb{Target Set Reconfiguration} problem,
given two target sets $X$ and $Y$ of the same size,
we are required to determine whether $X$ can be transformed into $Y$ by repeatedly swapping
one vertex in the current set with another vertex not in the current set
preserving every intermediate set as a target set.
In this paper, we investigate the complexity of \prb{Target Set Reconfiguration} in restricted cases.
On the hardness side,
we prove that \prb{Target Set Reconfiguration} is \PSPACE-complete on
bipartite planar graphs of degree $3$ and $4$ and of threshold $2$,
bipartite $3$-regular graphs and planar $3$-regular graphs of threshold $1$ and $2$, and split graphs, which is in contrast to the fact that
a special case called \prb{Vertex Cover Reconfiguration} is in \cP for the last graph class.
On the positive side, we present a polynomial-time algorithm for
\prb{Target Set Reconfiguration} on
graphs of maximum degree $2$ and trees.
The latter result can be thought of as a generalization of that for \prb{Vertex Cover Reconfiguration}.
\end{abstract}

\section{Introduction}
\emph{Combinatorial reconfiguration} is a research field studying the following problem:
Given a pair of feasible solutions for a search problem, can we find a step-by-step transformation between them while keeping the feasibility?
Studying such problems may help understand the structure of the solution space of a search problem and have applications in dynamic and changing environments \cite{mouawad2015reconfiguration,haddadan2016complexity}.
Countless reconfiguration problems are derived from classical search problems, e.g.,
\prb{Boolean Satisfiability} \cite{gopalan2009connectivity,makino2011exact,mouawad2017shortest},
\prb{Clique}, \prb{Matching} \cite{ito2011complexity},
\prb{Coloring} \cite{cereceda2008connectedness,bonsma2009finding,cereceda2011finding},
\prb{Subset Sum} \cite{ito2014approximability}, and
\prb{Shortest Path} \cite{kaminski2011shortest,bonsma2013complexity}.
See also the survey of Nishimura \cite{nishimura2018introduction} and
van den Heuvel \cite{heuvel13complexity}.
One of the most well-studied reconfiguration problems is
based on \prb{Vertex Cover} \cite{hearn2005pspace,ito2011complexity,bonsma2016independent,bonsma2014reconfiguring,kaminski2012complexity,lokshtanov2019complexity,wrochn2018reconfiguration}.
Given a graph $G$ and two vertex covers $X$ and $Y$ of $G$,\footnote{A \emph{vertex cover} of $G$ is a vertex set that includes at least one endpoint of every edge.}
\prb{Vertex Cover Reconfiguration} requests to decide if
$X$ can be transformed into $Y$ by applying a sequence of prespecified transformation rules preserving every intermediate set as a vertex cover.
Such a sequence is called a \emph{reconfiguration sequence}.
Under a \emph{token jumping rule} \cite{kaminski2012complexity},
we can add one vertex and remove another vertex simultaneously by a single step 
(see \cref{sec:pre} for a formal definition), where
\prb{Vertex Cover Reconfiguration} was shown to be \PSPACE-complete \cite{hearn2005pspace,ito2011complexity,kaminski2012complexity}.

In this paper, we present an initial study on reconfigurability of \emph{target sets} (to the best of our knowledge).
\prb{Target Set Selection} is a combinatorial optimization problem on a graph
that finds applications in social network analysis \cite{chen2009approximability,kempe2003maximizing} and distributed computing \cite{peleg2002local,peleg1998size}.
Given a graph $G$ with vertex thresholds $\tau$,
we consider a dynamic process where
vertex $v$ becomes activated once at least $\tau(v)$ of $v$'s neighbors are activated,
which models the spread of influence, information, and opinion over a network.
A vertex set $S$ is called a \emph{target set} if all vertices of $G$ would be activated when initially activating vertices of $S$.
The objective of \prb{Target Set Selection} is to identify the minimum target set of $G$.
This problem generalizes well-studied \prb{Vertex Cover} and \prb{Feedback Vertex Set} problems:
It is known \cite{dreyer2000applications,dreyer2009irreversible} that
a target set is a vertex cover (resp.~a feedback vertex set)
if $\tau(v)$ is the degree of $v$ (resp.~the degree of $v$ minus $1$) for every vertex $v$.
In a reconfiguration version of \prb{Target Set Selection}, namely, \prb{Target Set Reconfiguration},
we are asked to decide if there exists a reconfiguration sequence between two particular target sets.

\subsection{Related Work and Known Results}

\subparagraph{Reconfiguration Problems.}
We review known results for \prb{Vertex Cover Reconfiguration} (\prb{VC-R} for short) and
\prb{Feedback Vertex Set Reconfiguration} (\prb{FVS-R} for short),
which are included as a special case of \prb{Target Set Reconfiguration}.
Hearn and Demaine~\cite{hearn2005pspace} are the first to prove that
\prb{VC-R} is \PSPACE-complete on planar graphs of maximum degree $3$
by reducing from \prb{Nondeterministic Constraint Logic} (see also \cite{hearn2009games}).
Since the unified framework of reconfiguration has been established by
{Ito, Demaine, Harvey, Papadimitriou, Sideri, Uehara, and Uno}~\cite{ito2011complexity},
great effort has been devoted to analyzing the restricted-case hardness and solvability of \prb{VC-R}, summarized in \cref{tab:class}.
Typically, a search problem in \cP induces a reconfiguration counterpart that belongs to \cP;
e.g., polynomial-time algorithms are known for
chordal graphs, split graphs, trees \cite{kaminski2012complexity,ito2016reconfiguration,mouawad2018vertex},
claw-free graphs \cite{bonsma2014reconfiguring},
and cacti \cite{mouawad2018vertex}.
Some exceptions, however, are known:
\prb{VC-R} is
\PSPACE-complete on perfect graphs \cite{kaminski2012complexity} and bounded-treewidth graphs \cite{wrochn2018reconfiguration}, and
it is \NP-complete on bipartite graphs \cite{lokshtanov2019complexity},
for which \prb{Vertex Cover} is in \cP.
Since any hardness result of \prb{VC-R}
directly applies to \prb{Target Set Reconfiguration},
it is interesting to explore
the complexity of \prb{Target Set Reconfiguration}
for cases where \prb{VC-R} is in \cP.

Some parameterized complexity results \cite{downey2012parameterized,cygan2015parameterized} for \prb{FVS-R} are known:
\prb{FVS-R} is fixed-parameter tractable when parameterized by the size of a feedback vertex set, but
it is \Wone-hard when parameterized by the length of reconfiguration sequences \cite{mouawad2014reconfiguration}.
On the other hand,
parameterization by the length of reconfiguration sequences and the treewidth of a graph is fixed-parameter tractable \cite{mouawad2017parameterized}.
Ito and Otachi~\cite{ito2019reconfiguration} showed that
\prb{FVS-R} is polynomial-time solvable on split graphs and interval graphs.

\begin{table}[tbp]
    \centering
    \caption{Complexity of \prb{Target Set Selection} (\prb{TSS}), \prb{Target Set Reconfiguration} (\prb{TS-R}), \prb{Vertex Cover} (\prb{VC}), and \prb{Vertex Cover Reconfiguration} (\prb{VC-R}) on restricted graph classes.
    }
    \label{tab:class}
    \footnotesize
    \begin{tabular}{c|l|l|l|l}
    \toprule
    graph class & \prb{TSS} & \prb{TS-R} & \prb{VC} & \prb{VC-R} \\
    \midrule
    planar & \NP-c (from \prb{VC}) & \PSPACE-c (from \prb{VC-R}) & \NP-c & \PSPACE-c \cite{hearn2005pspace,ito2011complexity,kaminski2012complexity} \\
    even-hole-free & \NP-c (from split) & \PSPACE-c (from split) & open & \cP \cite{kaminski2012complexity,ito2016reconfiguration,mouawad2018vertex} \\
    perfect & \NP-c (from split) & \PSPACE-c (from \prb{VC-R}) & \cP & \PSPACE-c \cite{kaminski2012complexity} \\
    chordal & \NP-c (from split) & \PSPACE-c (from split) & \cP & \cP (since even-hole-free) \\
    split & \NP-c \cite{nichterlein2013tractable} & \PSPACE-c (\cref{thm:split}) & \cP & \cP (since even-hole-free) \\
    claw-free & \NP-c \cite{munaro2017line} & \cellcolor{Red3!20}open & \cP & \cP \cite{bonsma2014reconfiguring} \\
    tree & \cP \cite{chen2009approximability} & \cP (\cref{thm:tree}) & \cP & \cP (since even-hole-free) \\
    bipartite & \NP-c \cite{chen2009approximability} & \PSPACE-c (\cref{thm:pb342}) & \cP & \NP-c \cite{lokshtanov2019complexity} \\
    bounded treewidth & \cP \cite{ben-zwi2011treewidth} & \PSPACE-c (from \prb{VC-R}) & \cP & \PSPACE-c \cite{wrochn2018reconfiguration} \\
    cactus & \cP \cite{ben-zwi2011treewidth,chiang2013some} & \cellcolor{Red3!20}open & \cP & \cP \cite{mouawad2018vertex} \\
    \bottomrule
    \end{tabular}
\end{table}

\subparagraph{Target Set Selection.}
Since \prb{Target Set Selection} (\prb{TSS} for short) naturally arises in many different fields, it is also known by various names such as
\emph{irreversible $k$-conversion sets} \cite{dreyer2009irreversible,centeno2011irreversible} and
\emph{dynamic monopolies} \cite{peleg1998size,peleg2002local}.
We review the complexity results of \prb{TSS} in restricted cases.
One direction is to investigate the case of bounded degree and/or bounded threshold.
The following settings of vertex thresholds are well established:
(1) \emph{majority thresholds}, where a vertex becomes activated if at least half of its neighbors are already activated;
(2) \emph{constant thresholds}, that is,
all thresholds are some constant $t$, e.g., $t=2$.
Peleg~\cite{peleg2002local} showed that it is \NP-hard to find a minimum target set for majority thresholds.
Dreyer~\cite{dreyer2000applications} and Dreyer and Roberts~\cite{dreyer2009irreversible} proved that a target set
is a vertex cover if $\tau(v)$ is the degree of $v$ for every vertex $v$, and
it is a feedback vertex set if $\tau(v)$ is the degree of $v$ minus $1$.
So, \prb{TSS} turns out to be \NP-hard even when
the threshold $\tau(v)$ of every vertex $v$ is a constant $t$ for any $t \geq 3$.
Chen \cite{chen2009approximability} provided the first \NP-hardness result for the case of $t=2$,
which is tight in the sense that the case of $t=1$ is trivially solved.
In fact, Chen~\cite{chen2009approximability} gave \NP-hardness of approximating \prb{TSS} within a polylogarithmic factor.
Subsequently, \NP-hardness under $t = 2$ was established for
graphs of maximum degree $11$ \cite{centeno2011irreversible} and graphs of maximum degree $4$ \cite{penso2014p3,kyncl2017irreversible}.
On cubic (i.e., $3$-regular) graphs of threshold $2$,
\prb{TSS} is equivalent to \prb{Feedback Vertex Set},
which is solvable in polynomial time \cite{ueno1988nonseparating}.
Further, polynomial-time algorithms for \prb{TSS} on subcubic graphs of threshold $2$ are known \cite{takaoka2015note,kyncl2017irreversible}.
Feige and Kogan~\cite{feige2019target} reported hardness-of-approximation results  of \prb{TSS} for several classes of bounded-degree graphs, including
$3$-regular graphs of threshold $1$ and $2$,
$4$-regular graphs of threshold $2$, and
$4$-regular graphs of threshold $3$.
Note that the case of maximum degree $2$ is trivial \cite{dreyer2009irreversible}.

A different direction is to consider restricted classes of graphs.
Chen \cite{chen2009approximability} gave a linear-time algorithm for trees.
Ben-Zwi, Hermelin, Lokshtanov, and Newman~\cite{ben-zwi2011treewidth} developed an $n^{O(\mathrm{tw})}$-time algorithm,
where $n$ is the number of vertices, and $\mathrm{tw}$ is the treewidth of a graph,
as a generalization of Chen~\cite{chen2009approximability}'s algorithm, and
they ruled out the existence of an $n^{o(\sqrt{\mathrm{tw}})}$-time algorithm under some plausible complexity-theoretic assumption.
Other graph classes rendering \prb{TSS} tractable include
block-cactus graphs \cite{chiang2013some},
cliques \cite{nichterlein2013tractable}, 
chordal graphs with bounded thresholds \cite{chiang2013some}, and
interval graphs with bounded thresholds \cite{bessy2019dynamic}.
Conversely, \NP-hardness was shown for
split graphs \cite{nichterlein2013tractable},
claw-free graphs \cite{munaro2017line},
planar graphs \cite{dreyer2009irreversible}, and
bipartite graphs \cite{dreyer2009irreversible}.
Parameterized complexity of \prb{TSS} is examined for numerous parameters \cite{nichterlein2013tractable,chopin2014constant,dvorak2018target,bazgan2014parameterized}.

\begin{table}[tbp]
    \centering
    \caption{Complexity of \prb{Target Set Reconfiguration} on small degree/threshold graphs.}
    \label{tab:degree}
    \small
    \begin{tabular}{|c|l|c|c|c|c|}
    \hhline{~~----}
    \multicolumn{2}{c|}{} & \multicolumn{4}{c|}{\textbf{vertex degree}} \\ \hhline{~~----}
    \multicolumn{2}{c|}{} & $d(v)\in\{1,2\}$ & $d(v)\in\{2,3\}$ & $d(v)= 3$ & $d(v)\in\{3,4\}$ \\
    \hline
    \multirow{4}{*}{\begin{rotatebox}{90}{\textbf{threshold}}\end{rotatebox}} & $\tau(v)=1$ & \multicolumn{4}{c|}{\cP (\cref{obs:thr-1})} \\ \hhline{~-----}
    &$\tau(v) \in \{1,2\}$ & \cP (\cref{thm:degree-2}) & \multicolumn{2}{c|}{\PSPACE-c (\cref{thm:b312})} & \PSPACE-c \\ \hhline{~----~}
    & $\tau(v)=2$ &  & \cellcolor{Red3!20}open & \cellcolor{Red3!20}open & (\cref{thm:pb342})  \\ \hhline{~-~---}
    & $\tau(v)=3$ & \multicolumn{2}{c|}{} & \multicolumn{2}{c|}{\PSPACE-c (\cite{hearn2005pspace,ito2011complexity,kaminski2012complexity}; \cref{thm:33})}  \\ \hhline{--~~--}
    \end{tabular}
\end{table}

\subsection{Our Results}
In this paper, we study the complexity of
\prb{Target Set Reconfiguration} (\prb{TS-R} for short) in restricted cases,
aiming to reveal a dividing line between easy and hard instances.
\cref{sec:degree} examines \emph{small-degree graphs},
the results for which are outlined in \cref{tab:degree}.
One of the simplest cases is when all thresholds are $1$, which
ensures reconfigurability between any pair of target sets (\cref{obs:thr-1}).
Graphs of maximum degree $2$ are seemingly easy to handle
since they consist only of paths and cycles.
However, there exist a nontrivial pair of reconfigurable target sets, requiring a kind of ``detour'' (see \cref{subsec:degree-2}).
We devise a characterization of reconfigurable target sets by careful analysis,
yielding a polynomial-time algorithm (\cref{thm:degree-2}).
Once a graph can include degree-$3$ vertices,
\prb{TS-R} becomes computationally challenging.
We first show \PSPACE-completeness on bipartite planar graphs of degree $3$ and $4$ and of threshold $2$ (\cref{thm:pb342}).
This restricted-case result is of particular interest because
it satisfies constant and majority thresholds simultaneously.
On cubic graphs,
the case of $t=3$ is identical to \prb{VC-R},
which is known to be \PSPACE-complete (see \cite{hearn2005pspace,ito2011complexity,kaminski2012complexity} and \cref{thm:33}).
Besides, we derive \PSPACE-completeness on bipartite cubic graphs and planar cubic graphs even if
thresholds are taken from $\{1,2\}$ (\cref{thm:b312}).
Our proofs involve several gadgets that are constructed so carefully that they preserve reconfigurability.

\cref{sec:class} explores \emph{restricted graph classes}, the results for which are summarized in \cref{tab:class}.
On the algorithmic side, we develop a polynomial-time algorithm for \prb{TS-R} on trees (\cref{thm:tree}), which can be thought of as a generalization of that for \prb{VC-R} \cite{kaminski2012complexity,ito2016reconfiguration,mouawad2018vertex}.
Similar to the case of other reconfiguration problems on trees \cite{haddadan2016complexity,demaine2015linear},
we demonstrate that
any pair of target sets is reconfigurable, and that
an actual reconfiguration sequence can be found in polynomial time.
On the hardness side, we prove that \prb{TS-R} is \PSPACE-complete on
split graphs (\cref{thm:split}),
on which \prb{VC-R} belongs to \cP.
This result relies on a technique for reducing from \prb{Hitting Set} by
{Nichterlein, Niedermeier, Uhlmann, and Weller}~\cite{nichterlein2013tractable}.

Proofs of the statements marked with ``$\star$'' are deferred to \cref{app:proofs}.

\section{Preliminaries}
\label{sec:pre}

\subsection{Notations and Definitions}
For nonnegative integers $m$ and $n$ with $m \leq n$,
we define $ [n] \triangleq \{1, 2, \ldots, n\} $ and
$[m\isep n] \triangleq \{m,m+1,\ldots, n-1,n\}$.
A \emph{sequence} $\calS$ consisting of sets $S^{(0)}, S^{(1)}, \ldots, S^{(\ell)}$
is denoted as $\langle S^{(0)}, S^{(1)}, \ldots, S^{(\ell)} \rangle$, and we write $S^{(i)} \in \calS$ to mean that $S^{(i)}$ appears in $\calS$ (at least once).
The symbol $\uplus$ is used to emphasize that the union is taken over two \emph{disjoint} sets.
For a graph $G=(V,E)$,
let $V(G)$ and $E(G)$ denote the vertex set $V$ and the edge set $E$ of $G$, respectively.
We assume that graphs are simple; i.e., they have no self-loops or multi-edges.   
For a vertex $v$ of $G$,
we denote the neighborhood of $v$ by $N_G(v) \triangleq \{ u \mid (u,v) \in E\}$ and
the degree of $v$ by $d_G(v) \triangleq |N_G(v)|$.
We omit the subscript when $G$ is clear from the context.
For a vertex set $S \subseteq V$,
we write $G[S]$ for denoting the subgraph of $G$ induced by $S$, and
we write $G - S$ for denoting the induced subgraph $G[V \setminus S]$.
In this paper,
a \emph{threshold} function $\tau\colon V \to \bbN$ is often associated with graph $G=(V,E)$.
Hence, we also refer to a triplet $G=(V,E,\tau)$ as a graph.
A vertex $v$ of $G$ is referred to as a \emph{$(d',\tau')$-vertex} if $d(v) = d'$ and $\tau(v) = \tau'$.
A graph $G=(V,E,\tau)$ is referred to as a \emph{$(D,T)$-graph} for two integer sets $D$ and $T$ if 
$d(v) \in D$ and $\tau(v) \in T$ for all $v \in V$.

We define the \emph{activation process} over a graph $G=(V,E,\tau)$.
Each vertex takes either of two states: \emph{active} or \emph{inactive}.
For a \emph{seed set} $S \subseteq V$,
we define $\calA_G^{(t)}(S)$ as the set of already activated vertices at discrete-time step $t$.
Initially, the vertices of $S$ are active, and the others are inactive;
i.e., $\calA_G^{(0)}(S) \triangleq S$.
Given $\calA_G^{(t-1)}(S)$ at step $t-1$,
we verify whether each inactive vertex $v$ has at least $\tau(v)$ active neighbors.
If this is the case,
then $v$ becomes active at step $t$; i.e., $v$ is added into $\calA_G^{(t)}(S)$.
This process is \emph{irreversible}; i.e.,
an active vertex may not become inactive.
Formally, the set of active vertices at step $t \geq 1$ is recursively defined as:
\begin{align}
\calA_G^{(t)}(S) \triangleq \calA_G^{(t-1)}(S) \cup \left\{ v \in V \mid \left|N_G(v) \cap \calA_G^{(t-1)}(S)\right| \geq \tau(v) \right\}.
\end{align}
Observe that $\calA_G^{(n)}(S) = \calA_G^{(n+1)}(S)$ for $n \triangleq |V|$
by the fact that $\calA_G^{(t-1)}(S) \subseteq \calA_G^{(t)}(S)$ for all $t \geq 1$.
Therefore, we define the active vertex set of $S$ as $\calA_G(S) \triangleq \calA_G^{(n)}(S)$, and
we say that $S$ \emph{activates} a vertex $v$ or $v$ is \emph{activated by} $S$ in $G$ if $v \in \calA_G(S)$.
In particular, if $S$ activates the whole graph; i.e., $\calA_G(S) = V$, $S$ is called a \emph{target set} of $G$.
The \prb{Target Set Selection} problem is defined as follows.
\begin{problem}[\prb{Target Set Selection}]
Given a graph $G=(V,E,\tau)$,
find a minimum target set of $G$.
\end{problem}
Throughout this paper,
we assume that $1 \leq \tau(v) \leq d(v)$ for all $v \in V(G)$.
This is because if $\tau(v) > d(v)$, then any target set of $G$ must include $v$;
if $\tau(v)=0$, then $v$ is not included in any minimum target set of $G$ \cite[Observation 1]{nichterlein2013tractable}.\footnote{
Note that this assumption forces a graph to have no isolated vertex.}

We then formulate a reconfiguration version of \prb{Target Set Selection}
according to the reconfiguration framework of
{Ito, Demaine, Harvey, Papadimitriou, Sideri, Uehara, and Uno}~\cite{ito2011complexity}.
We consider the following two types of \emph{reconfiguration steps},
which specify how a target set can be transformed.
\begin{description}
    \item[\emph{Token jumping} (\TJ)] \cite{kaminski2012complexity}:
    Given a target set,
    a \TJ-step can remove one vertex from it \emph{and} add another vertex not in it simultaneously, which does not change the set size.
    \item[\emph{Token addition or removal} (\TAR)] \cite{ito2011complexity}:
    Given a target set, a \TAR-step can 
    remove a vertex from it \emph{or} add a vertex not in it. 
\end{description}
For two target sets $X$ and $Y$,
a \emph{reconfiguration sequence from $X$ to $Y$}
is a sequence of target sets
$\calS = \langle S^{(0)}, S^{(1)}, \ldots, S^{(\ell)} \rangle $
starting from $X$ (i.e., $S^{(0)}=X$) and ending with $Y$ (i.e., $S^{(\ell)} = Y$) such that
$S^{(i)}$ is obtained from $S^{(i-1)}$ by a single reconfiguration step for $i \in [\ell]$.
The \emph{length} $\ell$ of $\calS$ is defined as the number of sets in it minus $1$.
If $\calS$ consists only of \TJ-steps, then it is called a \emph{\TJ-sequence};
if $\calS$ consists only of \TAR-steps and every set in $\calS$ is of size at most $k+1$, then it is called a \emph{$k$-\TAR-sequence}.
Moreover,
we say that $X$ and $Y$ are \emph{\TJ-reconfigurable} on $G$
if there exists a \TJ-sequence of target sets of $G$ from $X$ to $Y$;
we say that $X$ and $Y$ are \emph{$k$-\TAR-reconfigurable} on $G$
if there exists a $k$-\TAR-sequence from $X$ to $Y$.
We define the \prb{Target Set Reconfiguration} problem as follows.
\begin{problem}[\prb{Target Set Reconfiguration}]
Given a graph $G=(V,E,\tau)$ and two target sets $X$ and $Y$ of the same size,
decide if $X$ and $Y$ are \TJ-reconfigurable or not.
\end{problem}
Observe easily that this problem is in \PSPACE \cite{heuvel13complexity}.
Note that the present problem definition does not request an actual \TJ-sequence.
We concern \TJ-reconfigurability only because it is essentially equivalent to $k$-\TAR-reconfigurability.
We define \prb{Minimum Target Set Reconfiguration} as a special case where $X$, $Y$, and all intermediate sets are promised to be minimum.

\subsection{Useful Lemmas}
Here, we introduce some lemmas, which are convenient for proving our results in the subsequent sections.
We first define \TJN-sequences that we use for a technical reason.
Given a target set, a \emph{\TJN-step} can perform either a \TJ-step \emph{or} do nothing.
A \emph{\TJN-sequence} is a reconfiguration sequence consisting only of \TJN-steps.
We say that $X$ and $Y$ are \emph{\TJN-reconfigurable}
if there exists a \TJN-sequence from $X$ to $Y$.
The following is trivial by definition.

\begin{observation}
\label{obs:tjn}
Let $X$ and $Y$ be two target sets of a graph $G$.
Then, $X$ and $Y$ are \TJ-reconfigurable on $G$ if and only if
they are \TJN-reconfigurable on $G$.
\end{observation}

We then show the equivalence between \TJ-reconfigurability and \TAR-reconfigurability, whose proof is an adaptation of
{Kami{\'n}ski, Medvedev, and Milani{\v{c}}}~\cite[Theorem 1]{kaminski2012complexity}.

\begin{observation}
\label{obs:TJ-TAR}
Let $G$ be a graph and $X$ and $Y$ be two size-$k$ target sets of $G$.
Then, $X$ and $Y$ are \TJ-reconfigurable if and only if
they are $k$-\TAR-reconfigurable.
\end{observation}
\begin{proof}
Suppose that we are given a \TJ-sequence $\langle S^{(i)} \rangle_{i \in [0\isep\ell]}$  from $X$ to $Y$.
Observe that a single \TJ-step from $S^{(i-1)}$ to $S^{(i)}$ can be converted into two \TAR-steps:
add a vertex $S^{(i)} \setminus S^{(i-1)}$ to $S^{(i-1)}$ to obtain $S^{(i-1)} \cup S^{(i)}$, and
remove a vertex $S^{(i-1)} \setminus S^{(i)}$ from $S^{(i-1)} \cup S^{(i)}$ to obtain $S^{(i)}$.
The resulting sequence is a $k$-\TAR-sequence from $X$ to $Y$.
This completes the only-if direction.

Suppose then that we are given a $k$-\TAR-sequence $\langle S^{(i)} \rangle_{i \in [0\isep\ell]}$ from $X$ to $Y$.
Until we obtain a $k$-\TAR-sequence consisting of target sets of size $k$ or $k+1$,
we modify the current $k$-\TAR-sequence according to the following procedure:
Let $\sigma$ be a length-$2$ subsequence of the current sequence consisting of
removal of vertex $x$ and addition of vertex $y$ such that
the middle set is of size less than $k$, say,
$\sigma = \langle S, S \setminus \{x\}, S \setminus \{x\} \cup \{y\} \rangle$, where $|S \setminus \{x\}| \leq k-1$.
If $x=y$, then
we can remove the two consecutive sets $S \setminus \{x\}$ and $S \setminus \{x\} \cup \{y\}$ to shorten the $k$-\TAR-sequence.
Otherwise, we can replace $\sigma$ with
the subsequence $\sigma'$ in which we first add $y$ and then remove $x$,
say, $\sigma' = \langle S, S \cup \{y\}, S \cup \{y\} \setminus \{x\} \rangle$,
which is still a $k$-\TAR-sequence.
One can continue this procedure until the resulting $k$-\TAR-sequence consists of target sets of size $k$ or $k+1$,
which turns out to be a \TJ-sequence, completing the proof.
\end{proof}

We introduce a combinatorial characterization of target sets due to Ackerman, Ben-Zwi, and Wolfovitz~\cite{ackerman2010combinatorial}.

\begin{theorem}[Adaptation of Lemma 2.1 of \cite{ackerman2010combinatorial}]
\label{thm:ackerman}
For a graph $G=(V,E,\tau)$,
a seed set $S\subseteq V$ is a target set of $G$ (i.e., $\calA_G(S) = V$) if and only if
there exists an acyclic orientation $D$ of $G$ such that
$d^-_D(v) \geq \tau(v)$ for every vertex $v \in V \setminus S$,
where $d^-_D(v)$ is the number of edges entering into $v$ in  $D$.
\end{theorem}

For a graph $G=(V,E,\tau)$ and a seed set $S \subseteq V$,
the \emph{residual} is defined as $G_S \triangleq (V_S,E_S,\tau_S)$, where
$V_S$ is the set of vertices that would not have been activated by $S$ on $G$; i.e., $V_S \triangleq V \setminus \calA_G(S)$,
$E_S \triangleq \{ (u,v) \in E \mid u \in V_S, v \in V_S \}$, and
$\tau_S(v)$ for each $v \in V_S$ is defined as $\tau(v)$ minus the number of $v$'s active neighbors; i.e.,
$\tau_S(v) \triangleq \tau(v) - |N_G(v) \cap \calA_G(S)|$.

\begin{lemma}
\label{lem:residual}
For a graph $G=(V,E,\tau)$ and two disjoint vertex subsets $S$ and $T$ of $V$,
let $G_S=(V_S,E_S,\tau_S)$ be the residual.
Then, $S \uplus T$ is a target set of $G$ if and only if $T \cap V_S$ is a target set of $G_S$.
Moreover, if $S \uplus T$ is a minimum target set of $G$,
then $T \cap V_S$ is a minimum target set of $G_S$.
\end{lemma}
\begin{proof}
We first prove the only-if direction.
Suppose that $S \uplus T$ is a target set of $G$. By \cref{thm:ackerman},
there exists an acyclic orientation $D$ of $G$ such that
$d^-_{D}(v) \geq \tau(v)$ for every vertex $v \in V \setminus (S \uplus T)$.
Define $D_S \triangleq D[V_S]$, which is an acyclic orientation of $G_S$.
By definition, it holds that
$\tau_S(v) = \tau(v) - |N_G(v) \cap \calA_G(S)|$ and
$d^-_{D_S}(v) = d^-_{D}(v) - |N^-_D(v) \cap \calA_G(S)|$ for every vertex $v \in V_S$.
Observing that $|N_G(v)| \geq |N^-_D(v)|$,
we have that $d^-_{D_S}(v) \geq \tau_S(v)$ for any $v \in V_S \setminus T$;
i.e., $T$ is a target set of $G_S$ by \cref{thm:ackerman}.

We then prove the if direction.
Suppose that $T \cap V_S$ is a target set of $G_S$.
By \cref{thm:ackerman},
there exists an acyclic orientation $D_S$ of $G_S$ such that
$d^-_{D_S}(v) \geq \tau_S(v)$ for every vertex $v \in V_S \setminus T$.
Since $S$ is a target set of $G[\calA_G(S)]$,
there exists an acyclic orientation $D'$ of $G[\calA_G(S)]$ such that $d^-_{D'}(v) \geq \tau(v)$ for every vertex $v \in \calA_G(S) \setminus S$.
Recall that $\calA_G(S) \uplus V_S = V$.
We define an orientation $D$ of $G$ as follows:
for each (undirected) edge $(u,v)$ of $G$,
\begin{description}
    \item[\textbf{(1)}] if $u,v \in \calA_G(S)$, then its direction coincides with that of $D'$;
    \item[\textbf{(2)}] if $u,v \in V_S$, then its direction coincides with that of $D_S$;
    \item[\textbf{(3)}] if $u \in \calA_G(S)$ and $v \in V_S$, it is directed from $u$ to $v$.
\end{description}
Note that $D$ is acyclic as well.
Then,
we have that $d^-_D(v) = d^-_{D'}(v) \geq \tau(v) $ for each vertex $v \in \calA_G(S) \setminus S$, and
we have that $d^-_D(v) = d^-_{D_S}(v) + |\calA_G(S) \cap N^-_D(v)| \geq \tau_S(v) + |\calA_G(S) \cap N_G(v)| = \tau(v) $ for each vertex $v \in V_S \setminus T$.
Accordingly, $S \uplus T$ is a target set of $G$ by \cref{thm:ackerman}.
The argument regarding minimality is obvious.
\end{proof}

The \emph{disjoint union} of two graphs $G_1$ and  $G_2$ is defined as
a graph $G_1 \oplus G_2$ with vertex set $V(G_1 \oplus G_2) \triangleq V(G_1) \uplus V(G_2)$ and edge set $E(G_1 \oplus G_2) \triangleq E(G_1) \uplus E(G_2)$.

\begin{lemma}
\label{lem:oplus}
Let $G_1$ and $G_2$ be two graphs, and
let $(X_1, Y_1)$ and $(X_2, Y_2)$ be pairs of two minimum target sets of $G_1$ and $G_2$,
respectively.
Then, $X_1 \uplus X_2$ and $Y_1 \uplus Y_2$ are \TJ-reconfigurable on the disjoint union $G_1 \oplus G_2$ if and only if
$X_1$ and $Y_1$ are \TJ-reconfigurable on $G_1$ and
$X_2$ and $Y_2$ are \TJ-reconfigurable on $G_2$.
\end{lemma}
\begin{proof}
Observe easily that
a seed set $S \subseteq V(G_1 \oplus G_2)$ is a minimum target set of $G_1 \oplus G_2$ if and only if
$S \cap V(G_1)$ is a minimum target set of $G_1$ and
$S \cap V(G_2)$ is a minimum target set of $G_2$.
Given
a \TJ-sequence $\calS_1 = \langle S_1^{(i)} \rangle_{i \in [0\isep\ell_1]}$ from $X_1$ to $Y_1$ for $G_1$ and
a \TJ-sequence $\calS_2 = \langle S_2^{(i)} \rangle_{i \in [0\isep\ell_2]}$ from $X_2$ to $Y_2$ for $G_2$,
we have that the sequence $\langle X_1 \uplus X_2, S_1^{(1)} \uplus X_2, \ldots, S_1^{(\ell_1-1)} \uplus X_2,
Y_1 \uplus X_2, Y_1 \uplus S_2^{(1)}, \ldots, Y_1 \uplus S_2^{(\ell_2-1)}, Y_1 \uplus Y_2 \rangle$
is a \TJ-sequence from $X_1 \uplus X_2$ to $Y_1 \uplus Y_2$ for $G_1 \oplus G_2$; i.e.,
$X_1 \uplus X_2$ and $Y_1 \uplus Y_2$ are \TJ-reconfigurable on $G_1 \oplus G_2$.
On the other hand, given a \TJ-sequence $\calS = \langle S^{(i)} \rangle_{i \in [0 \isep \ell]}$,
we have that
the sequence $\calS_1 \triangleq \langle S^{(i)} \cap V(G_1) \rangle_{i \in [0\isep \ell]}$ must be a \TJN-sequence from $X_1$ to $Y_1$ for $G_1$; i.e.,
$X_1$ and $Y_1$ are \TJ-reconfigurable on $G_1$, and
the sequence $\calS_2 = \langle S^{(i)} \cap V(G_2) \rangle_{i \in [0..\ell]}$ must be a \TJN-sequence from $X_2$ to $Y_2$ for $G_2$; i.e.,
$X_2$ and $Y_2$ are \TJ-reconfigurable on $G_2$, as desired.
\end{proof}

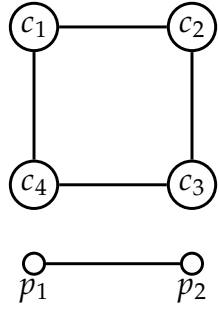
\begin{figure}[tbp]
    \centering
    \scalebox{0.7}{
    \begin{tikzpicture}
        \node[mynode] at (0,0) (c1){$c_1$};
        \node[mynode] at (3,0) (c2){$c_2$};
        \node[mynode] at (3,-3) (c3){$c_3$};
        \node[mynode] at (0,-3) (c4){$c_4$};
        \node[mymynode] at (0,-4.5) (p1){};
        \node[mymynode] at (3,-4.5) (p2){};
        \node[mylabel, below=0.5mm of p1, font=\LARGE]{$p_1$};
        \node[mylabel, below=0.5mm of p2, font=\LARGE]{$p_2$};
        \foreach \u / \v in {c1/c2, c2/c3, c3/c4, c4/c1, p1/p2}
            \draw[myedge] (\u)--(\v);
    \end{tikzpicture}
    }
    \caption{
        A graph $G$ of \cref{ex:oplus} consisting of 
        a length-$4$ cycle graph $G_1$ and a length-$1$ path graph $G_2$.
        $\bigcirc$ and $\circ$ represent threshold-$2$ and threshold-$1$ vertices, respectively.
    }
    \label{fig:oplus}
\end{figure}

Note that minimality is necessary for ensuring the only-if direction.
More precisely, we show by an example in \cref{fig:oplus} that
if either $(X_1,Y_1)$ or $(X_2,Y_2)$ is not minimum,
\TJ-reconfigurability between $X_1 \uplus X_2$ and $Y_1 \uplus Y_2$ in $G_1 \oplus G_2$
may not guarantee
\TJ-reconfigurability between $X_1$ and $Y_1$ in $G_1$ or $X_2$ and $Y_2$ in $G_2$.
\begin{example}
\label{ex:oplus}
Let $G_1$ be a cycle graph on four vertices $c_1,c_2,c_3,c_4$, and
let $G_2$ be a path graph on two vertices $p_1,p_2$.
Define vertex thresholds $\tau$ as
$\tau(c_i) = 2$ for all $c_i \in V(G_1)$ and
$\tau(p_i) = 1$ for all $p_i \in V(G_2)$.
Let
$X_1 = \{c_1,c_3\}$, $Y_1 = \{c_2,c_4\}$,
$X_2 = \{p_1,p_2\}$, and $Y_2 = \{p_1,p_2\}$.
Here, $X_1$ and $Y_1$ are minimum target sets of $G_1$, while 
$X_2$ and $Y_2$ are not minimum; i.e.,
$X_1 \uplus X_2$ and $Y_1 \uplus Y_2$ are not minimum for $G_1 \oplus G_2$.
Observe that $X_1 \uplus X_2$ and $Y_1 \uplus Y_2$ are \TJ-reconfigurable in $G_1 \oplus G_2$:
add $w_2$ and remove $v_1$;
add $w_4$ and remove $w_1$;
add $v_1$ and remove $w_3$.
However, $X_1$ and $Y_1$ are not \TJ-reconfigurable in $G_1$.
\end{example}

We then refer to \cite{nichterlein2013tractable} to show that 
a target set does not need to include threshold-$1$ vertices.
\begin{observation}[{Nichterlein et al.}~\cite{nichterlein2013tractable}]
\label{obs:replace}
For a graph $G=(V,E,\tau)$, let $S$ be a target set of $G$.
Let $v$ be a vertex in $S$ and $w $ be a neighbor of $v$ in $G$,
which may or may not be in $S$.
If $\tau(v) = 1$, then $S \setminus \{v\} \cup \{w\}$ is a target set of $G$.
\end{observation}

The \emph{subdivision} of an edge $(u,v)$ of graph $G$ consists of 
the removal of $(u,v)$ from $G$ and the addition of a new vertex $w$ and two edges $(u,w)$ and $(w,v)$.
Let $G'$ be a graph obtained from $G$ by 
subdividing an edge $(u,v) \in E(G)$ by a new vertex $w$ whose threshold is $1$.
For a seed set $S' \subseteq V'$ of $G'$,
we define $\phi_\mathrm{sd}(S') \subseteq V$ as
\begin{align}
    \phi_\mathrm{sd}(S') \triangleq
    \begin{cases}
    S' & \text{if } w \not\in S', \\
    S' \setminus \{w\} \cup \{u\} & \text{if } w \in S'.
    \end{cases}
\end{align}

\begin{lemma}
\label{lem:subdivision}
A seed set
$S' \subseteq V'$ is a minimum target set of $G'$
if and only if
$\phi_\mathrm{sd}(S')$ is a minimum target set of $G$.
Moreover,
two minimum target sets $X$ and $Y$ of $G$ are \TJ-reconfigurable on $G$ if and only if
they are \TJ-reconfigurable on $G'$.
\end{lemma}
\begin{proof}
One can verify that
if $S \subseteq V$ is a target set of of $G$, then
$S$ is also a target set of $G'$;
if $S' \subseteq V'$ is a target set of $G'$, then
$\phi_\mathrm{sd}(S')$ is a target set of $G$.
Since a minimum target set of $G'$ does not include $u$ and $w$ at the same time owing to \cref{obs:replace},
we have $|\phi_\mathrm{sd}(S')| = |S'|$ for a minimum target set $S'$ of $G'$, i.e.,
$\phi_\mathrm{sd}(S')$ is a minimum target set of $G$.

We then demonstrate that two minimum target sets $X$ and $Y$ of $G$ are \TJ-reconfigurable on $G$ if and only if they are \TJ-reconfigurable on $G'$.
Given a \TJ-sequence $\calS$ of minimum target sets of $G$ from $X$ to $Y$,
we find $\calS$ to be a \TJ-sequence of minimum target sets of $G'$ from $X$ to $Y$,
completing the only-if direction.
On the other hand,
given a \TJ-sequence $\calS'$ of minimum target sets of $G'$ from $X$ to $Y$,
we find the sequence
$ \calS = \langle \phi_\mathrm{sd}(S') \rangle_{S' \in \calS'} $
to be a \TJN-sequence of minimum target sets of $G$ from $\phi_\mathrm{sd}(X)=X$ to $\phi_\mathrm{sd}(Y)=Y$; i.e.,
$X$ and $Y$ are \TJ-reconfigurable on $G$,
completing the if direction.
\end{proof}

We finally introduce a \emph{one-way gadget} \cite{kyncl2017irreversible,charikar2016approximating,bazgan2014parameterizeda}, which
is defined as a graph $D=(V,E,\tau)$ such that
$V \triangleq \{t,h,b_1,b_2\}$,
$E \triangleq \{(t,b_1), (t,b_2), (h,b_1), (h,b_2)\}$, and
$\tau(t)=\tau(b_1)=\tau(b_2)=1, \tau(h)=2$.
We say that $D$ \emph{connects} from vertex $v$ to vertex $w$ if there exist two edges $(v,t)$ and $(w,h)$.
The vertices of $V$ are called the \emph{internal vertices} of a one-way gadget $D$.
Observe that $v$ activates vertices $t,h,b_1,b_2$, but $w$ does not.

\section{Small Degree Graphs}
\label{sec:degree}
In this section, we study the complexity of \prb{Target Set Reconfiguration}
on small degree graphs.
As a warm-up, we show that
threshold-$1$ graphs are amenable.

\begin{observation}
\label{obs:thr-1}
Let $G$ be a graph in which every vertex has threshold $1$.
Then, any two target sets of the same size are \TJ-reconfigurable.
\end{observation}
\begin{proof}
Suppose that $G$ consists of $c$ connected components, denoted $C_1, \ldots, C_c$.
Then, a minimum target set $S^*$ of $G$ includes exactly one vertex,
say, $v_i$, of each connected component $C_i$.
Hence, for any size-$k$ target set $S$, we can construct a $k$-\TAR-sequence from $S$ to $S^*$ as follows:
for each $i \in [c]$, add a vertex $v_i$ if $v_i \not\in S$ and remove the vertices of $C_i \setminus \{v_i\}$ one by one.
By \cref{obs:TJ-TAR}, any two target sets of the same size are \TJ-reconfigurable.
\end{proof}

\begin{figure}[tbp]
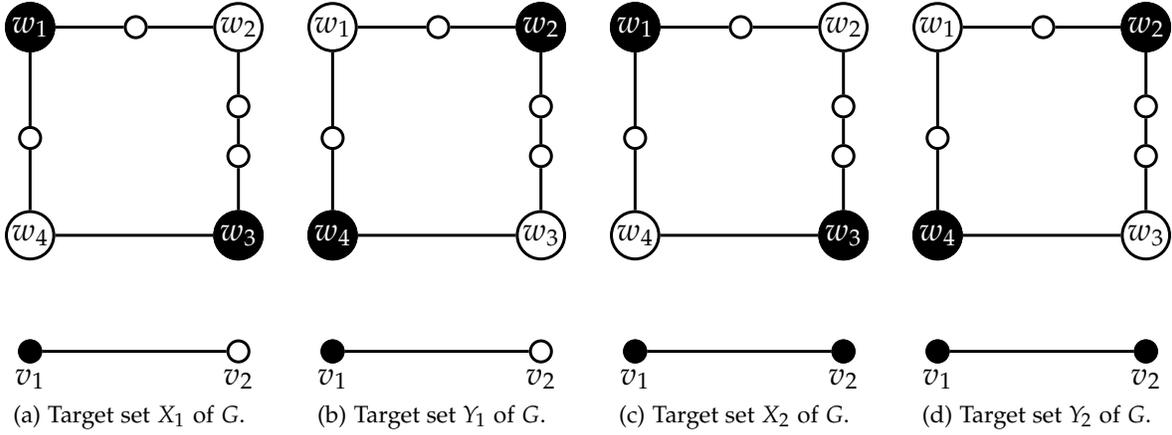

    \centering
    \null\hfill
    \subfloat[Target set $X_1$ of $G$. \label{fig:degree-2:X1}]{\scalebox{0.7}{
        \begin{tikzpicture}
        \input{fig_degree-2-temp}
        \draw(w1) node[mytoken]{$w_1$};
        \draw(w3) node[mytoken]{$w_3$};
        \draw(v1) node[mymynode, fill=black]{};
        \end{tikzpicture}
    }}
    \hfill
    \subfloat[Target set $Y_1$ of $G$. \label{fig:degree-2:Y1}]{\scalebox{0.7}{
        \begin{tikzpicture}
        \input{fig_degree-2-temp}

        \draw(w2) node[mytoken]{$w_2$};
        \draw(w4) node[mytoken]{$w_4$};
        \draw(v1) node[mymynode, fill=black]{};

        \end{tikzpicture}
    }}
    \hfill
    \subfloat[Target set $X_2$ of $G$. \label{fig:degree-2:X2}]{\scalebox{0.7}{
        \begin{tikzpicture}
        \input{fig_degree-2-temp}
        \draw(w1) node[mytoken]{$w_1$};
        \draw(w3) node[mytoken]{$w_3$};
        \draw(v1) node[mymynode, fill=black]{};
        \draw(v2) node[mymynode, fill=black]{};
        \end{tikzpicture}
    }}
    \hfill
    \subfloat[Target set $Y_2$ of $G$. \label{fig:degree-2:Y2}]{\scalebox{0.7}{
        \begin{tikzpicture}
        \input{fig_degree-2-temp}
        \draw(w2) node[mytoken]{$w_2$};
        \draw(w4) node[mytoken]{$w_4$};
        \draw(v1) node[mymynode, fill=black]{};
        \draw(v2) node[mymynode, fill=black]{};
        \end{tikzpicture}
    }}
    \hfill\null
    \caption{
        A graph $G$ of maximum degree $2$ and four target sets $X_1,Y_1,X_2,Y_2$.
        $\bigcirc$ and $\circ$ represent threshold-$2$ and threshold-$1$ vertices, respectively.
        Seed vertices are colored black $\bullet$.
        $X_1$ and $Y_1$ are not \TJ-reconfigurable while 
        $X_2$ and $Y_2$ are \TJ-reconfigurable.
    }
    \label{fig:degree-2}
\end{figure}

\subsection{Polynomial-time on Maximum Degree Two Graphs}
\label{subsec:degree-2}
We address a graph of maximum degree $2$.
Since such a graph consists of paths and cycles,
finding a minimum target set is an easy problem \cite{dreyer2009irreversible}.
On the other hand, the reconfiguration problem becomes somewhat intricate, as shown below.
\begin{example}
Take \cref{fig:degree-2} as an example of two similar but different instances of \prb{Target Set Reconfiguration}.
Here, a graph $G$ is made up of a cycle having four threshold-$2$ vertices $w_1,w_2,w_3,w_4$ and
a path consisting of two threshold-$1$ vertices $v_1,v_2$.
Two target sets $X_1$ and $Y_1$ respectively in \cref{fig:degree-2:X1,fig:degree-2:Y1}
are not \TJ-reconfigurable because any seed in $X_1$ cannot be moved.
On the other hand,
in \cref{fig:degree-2:X2,fig:degree-2:Y2},
we have the following \TJ-sequence from $X_2$ to $Y_2$:
add $c_2$ and remove $p_1$;
add $c_4$ and remove $c_1$;
add $p_1$ and remove $c_3$.
Note that this \TJ-sequence requires a kind of detour ($v_1$ appears in two \TJ-steps).
\end{example}
We show that \prb{Target Set Reconfiguration} on a graph of maximum degree $2$ is polynomial-time solvable, as stated below.

\begin{theorem}
\label{thm:degree-2} 
\prb{Target Set Reconfiguration} can be solved in polynomial time for graphs of maximum degree $2$.
Moreover, if the answer is ``yes,'' an actual \TJ-sequence can be found in polynomial time.
\end{theorem}

To prove \cref{thm:degree-2},
we characterize reconfigurable target sets on path and cycle graphs respectively in \cref{lem:path,lem:cycle}.

\begin{lemma}
\label{lem:path}
Let $G$ be a path graph including $m$ threshold-$2$ vertices.
Then, the size of the minimum target set is $\floor{\frac{m}{2}}+1$.
Any two target sets $X$ and $Y$ are $\max\{|X|, |Y|\}$-\TAR-reconfigurable.
In particular, when $|X|=|Y|$, they are \TJ-reconfigurable.
Moreover, an actual reconfiguration sequence can be found in polynomial time.
\end{lemma}

\begin{proof}
Of a path graph $G$,
let $w_1, \ldots, w_m$ denote $m$ threshold-$2$ vertices in a path order.
If $m=0$, the statement is obvious by \cref{obs:thr-1}.
Hereafter, suppose $m>0$.
Let $S$ be a minimum target set of $G$.
We can assume that $S$ only includes threshold-$2$ vertices,
because if $S$ includes threshold-$1$ vertices,
we can replace them by some threshold-$2$ vertices owing to \cref{obs:replace}.
Observe then that $S$ must include $w_i$ or $w_{i+1}$ for each $i \in [m-1]$;
thus, it holds that $|S| \geq \floor{\frac{m}{2}}+1$.
On the other hand,
the set $\{w_1, w_3, w_5, \ldots, w_{m-2}, w_m\}$ is a target set of $G$ if $m$ is positive odd, and
the set $\{w_1, w_3, w_5, \ldots, w_{m-1}, w_m\}$ if $m$ is positive even.
Therefore, the size of the minimum target set of $G$ is $\floor{\frac{m}{2}}+1$.

Since a path graph is a tree,
we use \cref{lem:tree-tar}, which will be proved later on and states that
any target set $S$ of a tree is $|S|$-\TAR-reconfigurable to some minimum target set.
Thus, any two target sets $X$ and $Y$ are $\max\{|X|, |Y|\}$-\TAR-reconfigurable,
which completes the proof.
\end{proof}

\begin{lemma}
\label{lem:cycle}
For a cycle graph $G$ including $m$ threshold-$2$ vertices,
we have the following:
\begin{itemize}
\item If $m = 0$: Any two target sets $X$ and $Y$ of $G$ are $\max\{|X|,|Y|\}$-\TAR-reconfigurable.
\item If $m$ is positive even:
The size of the minimum target set is $\frac{m}{2}$.
For a size-$k$ target set $S$ such that $k \geq \frac{m}{2}+1$ (i.e., $S$ is not minimum),
there exists a $k$-\TAR-sequence from $S$ to some minimum target set.
There are exactly two minimum target sets;
they are \TJ-reconfigurable when $m=2$, and
they are not \TJ-reconfigurable but $(\frac{m}{2}+1)$-\TAR-reconfigurable when $m \geq 4$.
\item If $m$ is positive odd:
The size of the minimum target set is $\frac{m+1}{2}$.
For a size-$k$ target set $S$ such that $k \geq \frac{m+1}{2}$,
there exists a $k$-\TAR-sequence from $S$ to
a special minimum target set consisting of threshold-$2$ vertices.
Moreover, such special minimum target sets are \TJ-reconfigurable to each other.
In particular, for $k \geq \frac{m+1}{2}$,
there exists a $k$-\TAR-sequence from a size-$k$ target set to some minimum target set.
\end{itemize}
Moreover, an actual reconfiguration sequence can be found in polynomial time.
\end{lemma}

\begin{proof}
Of a cycle graph $G$,
let $w_1, \ldots, w_m$ denote $m$ threshold-$2$ vertices in a cyclic order.
For notational convenience,
we assume that arithmetic operations regarding the subscript of
variables are performed over modulo $m$;
e.g., $w_{m+1} \equiv w_1$ and $w_{2m} \equiv w_{m}$.
In the case of $m=0$, the statement is obvious by \cref{obs:thr-1}.

\begin{figure}[tbp]
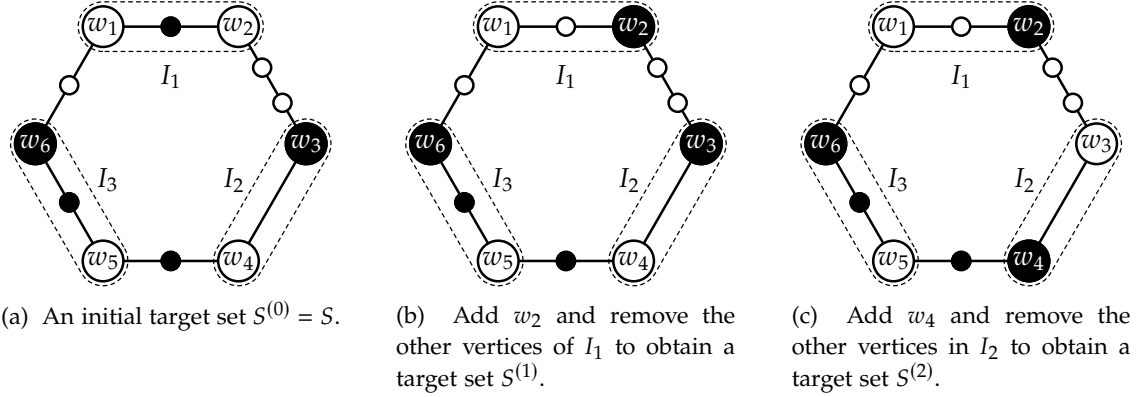
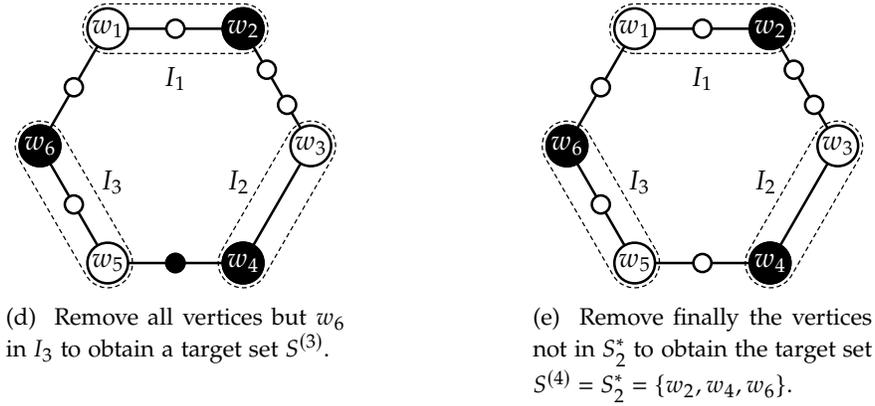

    \centering
    \null\hfill
    \subfloat[
        An initial target set $S^{(0)} = S$.
    ]{\scalebox{0.6}{
        \begin{tikzpicture}
        \input{fig_cycle-even-temp}
        \draw(w3) node[mytoken]{$w_3$};
        \draw(w6) node[mytoken]{$w_6$};
        \draw(a) node[mymynode, fill=black]{};
        \draw(d) node[mymynode, fill=black]{};
        \draw(e) node[mymynode, fill=black]{};
        \end{tikzpicture}
    }}
    \hfill
    \subfloat[
        Add $w_2$ and remove the other vertices of $I_1$ to obtain a target set $S^{(1)}$.
    ]{\scalebox{0.6}{
        \begin{tikzpicture}
        \input{fig_cycle-even-temp}
        \draw(w2) node[mytoken]{$w_2$};
        \draw(w3) node[mytoken]{$w_3$};
        \draw(w6) node[mytoken]{$w_6$};
        \draw(d) node[mymynode, fill=black]{};
        \draw(e) node[mymynode, fill=black]{};
        \end{tikzpicture}
    }}
    \hfill
    \subfloat[
        Add $w_4$ and remove the other vertices in $I_2$ to obtain a target set $S^{(2)}$.
    ]{\scalebox{0.6}{
        \begin{tikzpicture}
        \input{fig_cycle-even-temp}
        \draw(w2) node[mytoken]{$w_2$};
        \draw(w4) node[mytoken]{$w_4$};
        \draw(w6) node[mytoken]{$w_6$};
        \draw(d) node[mymynode, fill=black]{};
        \draw(e) node[mymynode, fill=black]{};
        \end{tikzpicture}
    }}
    \hfill\null
    \\
    \null\hfill
    \subfloat[
        Remove all vertices but $w_6$ in $I_3$ to obtain a target set $S^{(3)}$.
    ]{\scalebox{0.6}{
        \begin{tikzpicture}
        \input{fig_cycle-even-temp}
        \draw(w2) node[mytoken]{$w_2$};
        \draw(w4) node[mytoken]{$w_4$};
        \draw(w6) node[mytoken]{$w_6$};
        \draw(d) node[mymynode, fill=black]{};
        \end{tikzpicture}
    }}
    \hfill
    \subfloat[
        Remove finally the vertices not in $S^*_2$ to obtain the target set $S^{(4)} = S^*_2 = \{w_2,w_4,w_6\}$.
    ]{\scalebox{0.6}{
        \begin{tikzpicture}
        \input{fig_cycle-even-temp}
        \draw(w2) node[mytoken]{$w_2$};
        \draw(w4) node[mytoken]{$w_4$};
        \draw(w6) node[mytoken]{$w_6$};
        \end{tikzpicture}
    }}
    \hfill\null
    \caption{
        Illustration of \TAR-reconfigurability from an initial target set $S$ to $S^*_2$
        in a cycle graph including six threshold-$2$ vertices (\cref{lem:cycle}).
        $\bigcirc$ and $\circ$ represent threshold-$2$ and threshold-$1$ vertices, respectively.
        Seed vertices are colored black $\bullet$.
    }
    \label{fig:cycle-even}
\end{figure}

\medskip
\noindent
\textbf{Suppose that $m$ is positive even.}
We show that there exist two minimum target sets of size $\frac{m}{2}$.
We first claim that any minimum target set $S$ does not include threshold-$1$ vertices.
This is because if $S$ includes a threshold-$1$ vertex $v$,
the residual $G_{\{v\}}$ turns out to be a path graph having $m-2$ threshold-$2$ vertices,
whose minimum target set has size $\frac{m}{2}$ due to \cref{lem:path};
thus, $|S| \geq \frac{m}{2}+1$.
Observing further that any minimum target set must include $w_{i}$ or $w_{i+1}$ for all $i \in [m]$,
we come up with the following two minimum target sets:
$S^*_1 \triangleq \{w_1,w_3,\ldots,w_{m-1}\}$ and
$S^*_2 \triangleq \{w_2,w_4,\ldots,w_m\}$.

Let $S$ be a target set of size $k \geq \frac{m}{2} + 1$.
We construct a $k$-\TAR-sequence from $S$ to either $S^*_1$ or $S^*_2$.
For each $i \in [\frac{m}{2}]$, let $I_i$ denote
the unique path from $w_{2i-1}$ to $w_{2i}$ (including end vertices)
passing only through threshold-$1$ vertices.
Note that $\{I_1, \ldots, I_{\frac{m}{2}}\}$ forms a \emph{packing} of $V(G)$, and
any target set of $G$ includes at least one vertex of $I_i$ for all $i \in [\frac{m}{2}]$.
See \cref{fig:cycle-even} for an example.
Once $S$ contains $S^*_1$,
we just need to remove the vertices of $S \setminus S^*_1$ to obtain $S^*_1$; thus,
$S$ and $S^*_1$ are $k$-\TAR-reconfigurable.
Consider then that $S$ does not include some $w_i$ in $S^*_1$.
Here, we can safely assume that $w_1 \not\in S$ without loss of generality:
Otherwise, 
we can \emph{reorder} threshold-$2$ vertices to obtain $w'_1, \ldots, w'_m$
so that $w'_{j} = w_{j+i-1}$ for each $j \in [m]$.
Starting from $S^{(0)} \triangleq S$,
we transform $S^{(i-1)}$ into $S^{(i)}$ for each $i \in [\frac{m}{2}]$ by the following \TAR-steps (see \cref{fig:cycle-even}):
\begin{shadebox}
\begin{description}
\item[\textbf{Step 1.}] Add the vertex $w_{2i}$ to $S^{(i-1)}$ if $w_{2i} \not\in S^{(i-1)}$.
\item[\textbf{Step 2.}] Remove the vertices of $S^{(i-1)} \cap (I_i \setminus \{w_{2i}\})$ one by one.
\item[\textbf{Step 3.}] Let $S^{(i)}$ be the resulting set.
Note that $S^{(i)} \cap (\bigcup_{j \in [i]} I_j) = S^*_2 \cap (\bigcup_{j \in [i]} I_j)$.
\end{description}
\end{shadebox}
\noindent Since $S^*_2 \subseteq S^{(\frac{m}{2})}$,
we finally remove the vertices of
$S^{(\frac{m}{2})} \setminus S^*_2$ to obtain $S^{(\frac{m}{2}+1)} = S^*_2$.
Let $\calS$ be the resulting \TAR-sequence from $S$ to $S^*_2$.

We show that $S^{(i)}$ is a target set for each $i \in [0\isep\frac{m}{2}+1]$ by induction on $i$.
The base case of $i=0$ is obvious as $S^{(0)} = S$.
Suppose that $S^{(i-1)}$ is a target set for $i \in [\frac{m}{2}]$.
Consider the residual $G' = (V',E',\tau') \triangleq G_{S^{(i-1)} \setminus I_i}$.
Note that $S^{(i-1)} \setminus I_i = S^{(i)} \setminus I_i$.
Since $w_{2i-2} \in S^{(i-1)}$ by construction, $\tau'(w_{2i-1}) = 1$;
since $S^{(i-1)}$ (which is a target set by induction hypothesis) includes a vertex of $I_{i+1}$, $\tau'(w_{2i+1})=1$.
Thus, $w_{2i}$ is a unique vertex that may have threshold $2$ in $G'$.
Since $w_{2i} \in S^{(i)}$ by construction,
$(S^{(i)} \setminus I_i) \uplus \{w_{2i}\} = S^{(i)}$ must be a target set of $G$.
Of course, $S^{(\frac{m}{2}+1)} = S^*_2$ is a target set.

We then claim that $|S^{(i-1)}| \geq |S^{(i)}|$ for all $i \in [\frac{m}{2}]$.
If $w_{2i} \in S^{(i-1)}$, then the claim is obvious because we only remove vertices in Step 2.
Otherwise ($w_{2i} \not\in S^{(i-1)}$), we have
$S^{(i-1)} \cap I_i \neq \emptyset$;
thus, we remove at least one vertex in Step 2, implying that
$|S^{(i-1)}| \geq |S^{(i)}|$.
It is easy to see that $|S^{(\frac{m}{2})}| \geq |S^{(\frac{m}{2}+1)}|$.
Since any target set in the subsequence of $\calS$ from $S^{(i-1)}$ to $S^{(i)}$ 
has a size of at most $|S^{(i-1)}|+1$, the maximum size of any target set in $\calS$
from $S^{(0)} = S$ to $S^{(\frac{m}{2}+1)} = S^*_2$ is at most $k+1$;
i.e., $S$ and $S^*_2$ are $k$-\TAR-reconfigurable.

We then consider \TAR-reconfigurability between $S^*_1$ and $S^*_2$.
Consider transforming $S^*_1$ into $S^*_2$ by the following \TAR-steps:
\begin{shadebox}
\begin{description}
    \item[\textbf{Step 1.}] Add the vertex $w_m$ to $S^*_1$.
    \item[\textbf{Step 2.}] For each $i \in [\frac{m}{2}-1]$, add the vertex $w_{2i}$ and remove the vertex $w_{2i-1}$.
    \item[\textbf{Step 3.}] Remove the vertex $w_{m-1}$.
\end{description}
\end{shadebox}
\noindent Observe easily that
every intermediate vertex set is a target set of size at most $\frac{m}{2}+2$; i.e.,
$S^*_1$ and $S^*_2$ are $(\frac{m}{2}+1)$-\TAR reconfigurable.

Finally, when $m=2$, we have that $S^*_1 = \{w_1\}$ and $S^*_2=\{w_2\}$,
which are clearly \TJ-reconfigurable.
On the other hand, when $m\geq 4$,
$S^*_1$ and $S^*_2$ are not \TJ-reconfigurable as the symmetric difference between $S^*_1$ and $S^*_2$ has at least four vertices, as desired.

\begin{figure}[tbp]
    \centering
    \null\hfill
    \subfloat[
        An initial target set $S$.
    ]{\scalebox{0.6}{
        \begin{tikzpicture}
        \input{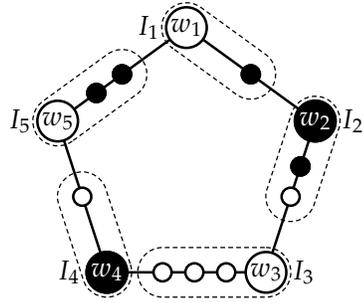}
        \foreach \u/\v in {w1/a, w2/c, w3/f, w4/g, w5/i}
            \draw[densely dashed, thick] \convexpath{\u,\v}{5.5mm};
        \draw(w2) node[mynode, fill=black]{\textcolor{white}{$w_2$}};
        \draw(w4) node[mynode, fill=black]{\textcolor{white}{$w_4$}};
        \draw(a) node[mymynode, fill=black]{};
        \draw(b) node[mymynode, fill=black]{};
        \draw(h) node[mymynode, fill=black]{};
        \draw(i) node[mymynode, fill=black]{};
        \end{tikzpicture}
    }}
    \hfill
    \subfloat[
        Add $w_1$ and remove the other vertices of $I_1$ to obtain a target set $S^{(0)}$.
    ]{\scalebox{0.6}{
        \begin{tikzpicture}
        \input{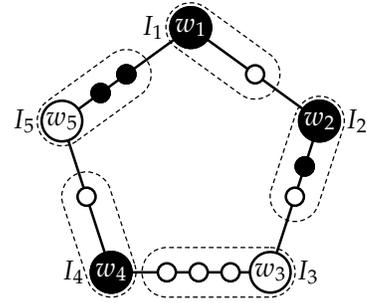}
        \foreach \u/\v in {w1/a, w2/c, w3/f, w4/g, w5/i}
            \draw[densely dashed, thick] \convexpath{\u,\v}{5.5mm};
        \draw(w1) node[mytoken]{$w_1$};
        \draw(w2) node[mytoken]{$w_2$};
        \draw(w4) node[mytoken]{$w_4$};
        \draw(b) node[mymynode, fill=black]{};
        \draw(h) node[mymynode, fill=black]{};
        \draw(i) node[mymynode, fill=black]{};
        \end{tikzpicture}
    }}
    \hfill\null
    \\
    \null\hfill
    \subfloat[
        Add $w_3$ and remove the other vertices of $I_2 \uplus I_3$ to obtain a target set $S^{(1)}$.
    ]{\scalebox{0.6}{
        \begin{tikzpicture}
        \input{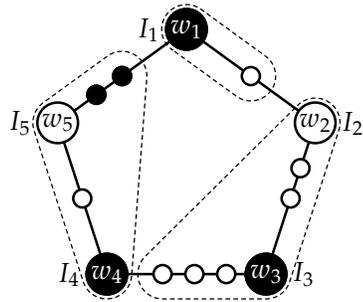}
        \foreach \u/\v in {w1/a}
            \draw[densely dashed, thick] \convexpath{\u,\v}{5.5mm};
        \draw[densely dashed, thick] \convexpath{w4,w5,i}{5.5mm};
        \draw[densely dashed, thick] \convexpath{w2,w3,f}{5.5mm};
        \draw(w1) node[mytoken]{$w_1$};
        \draw(w3) node[mytoken]{$w_3$};
        \draw(w4) node[mytoken]{$w_4$};
        \draw(h) node[mymynode, fill=black]{};
        \draw(i) node[mymynode, fill=black]{};
        \end{tikzpicture}
    }}
    \hfill
    \subfloat[
        Add $w_5$ and remove the other vertices of $I_4 \uplus I_5$ to obtain a minimum target set $S^{(2)}$ consisting only of threshold-$2$ vertices.
    ]{\scalebox{0.6}{
        \begin{tikzpicture}
        \input{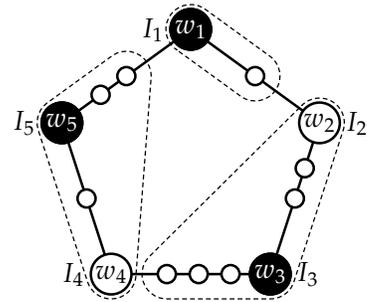}
        \foreach \u/\v in {w1/a}
            \draw[densely dashed, thick] \convexpath{\u,\v}{5.5mm};
        \draw[densely dashed, thick] \convexpath{w4,w5,i}{5.5mm};
        \draw[densely dashed, thick] \convexpath{w2,w3,f}{5.5mm};
        \draw(w1) node[mytoken]{$w_1$};
        \draw(w3) node[mytoken]{$w_3$};
        \draw(w5) node[mytoken]{$w_5$};
        \end{tikzpicture}
    }}
    \hfill\null
    \caption{
        Illustration of \TAR-reconfigurability from an initial target set $S$ to a minimum target set
        in a cycle graph including five threshold-$2$ vertices (\cref{lem:cycle}).
        $\bigcirc$ and $\circ$ represent threshold-$2$ and threshold-$1$ vertices, respectively.
        Seed vertices are colored black $\bullet$.
    }
    \label{fig:cycle-odd}
\end{figure}

\medskip
\noindent
\textbf{Suppose that $m$ is positive odd.}
For each $i \in [m]$, let $I_i$ denote
the unique path from $w_i$ to one vertex before $w_{i+1}$.
Note that $\{I_1, \ldots, I_m\}$ forms a \emph{partition} of $V(G)$.
See \cref{fig:cycle-odd} for an example.
Observe easily that for any target set $S$,
$|S \cap I_i| \geq 1$ or $|S \cap I_{i+1}| \geq 1$ for all $i \in [m]$.
It turns out that every target set $S$ satisfies that $|S| \geq \frac{m+1}{2}$.
On the other hand,
we can easily verify that the set $\{w_1, w_3, w_5, \ldots, w_{m-2}, w_m\}$ of size $\frac{m+1}{2}$ is a target set;
i.e., the size of the minimum target set is $\frac{m+1}{2}$.

For a target set $S$ of size $k$, we construct a $k$-\TAR-sequence $\calS$ from $S$ to a minimum target set consisting \emph{only} of threshold-$2$ vertices as follows (see \cref{fig:cycle-odd}):
\begin{shadebox}
\begin{description}
    \item[\textbf{Step 1.}] Let $i \in [m]$ be an integer such that $S \cap I_i \neq \emptyset$. Add the vertex $w_i$ to $S$ if $w_i \not\in S$ and remove the vertices of $S \cap (I_i \setminus \{w_i\})$ one by one.
    Let $S^{(0)}$ be the resulting set.
    \item[\textbf{Step 2.}] For each $j \in [\frac{m-1}{2}]$, do the following:
    \begin{description}
    \item[\textbf{Step 2-1.}] Add $w_{i+2j}$ if $w_{i+2j} \not\in S^{(j-1)}$.
    \item[\textbf{Step 2-2.}] Remove the vertices of $S^{(j-1)} \cap (I_{i+2j-1} \uplus I_{i+2j} \setminus \{w_{i+2j}\})$ one by one.
    \item[\textbf{Step 2-3.}] Let $S^{(j)}$ be the resulting set.
\end{description}
\end{description}
\end{shadebox}
\noindent
Similarly to the case of positive even $m$,
we can prove that
$S^{(i)}$ is a target set for all $i \in [0..\frac{m-1}{2}]$, and
prove that the maximum size of any target set in $\calS$ from $S$ to $S^{(\frac{m-1}{2})}$ is at most $k+1$.
Note that $S^{(\frac{m-1}{2})}$ consists of exactly $\frac{m+1}{2}$ threshold-2 vertices, which is a minimum target set.

We finally prove that any two minimum target sets consisting only of threshold-$2$ vertices are \TJ-reconfigurable.
For each $i \in [m]$, we define $S^*_i \triangleq \{ w_{i+2j} \mid j \in [0.. \frac{m-1}{2}] \}$.
It is not hard to see that any minimum target set consisting only of threshold-$2$ vertices is identical to $S^*_i$ for some $i \in [m]$.
Observe further that $S^*_{i}$ and $S^*_{i+2}$ for $i \in [m]$ are \TJ-reconfigurable: it suffices to add $w_{i+2}$ and remove $w_{i+1}$ by a single \TJ-step.
Since $m$ is an odd integer, we eventually have that $S^*_i$ and $S^*_j$ are \TJ-reconfigurable for any pair of $i, j \in[m]$, as desired.
\end{proof}

We are now ready to prove \cref{thm:degree-2}.
\begin{proof}[Proof of \cref{thm:degree-2}]
We say that a cycle graph is \emph{terrible} if the number of threshold-$2$ vertices in it is four or more and an even number.
Given a graph $G$ of maximum degree $2$ and two size-$k$ target sets $X$ and $Y$,
we demonstrate by case analysis that $X$ and $Y$ are not \TJ-reconfigurable
\emph{if and only if} the following conditions hold:
\begin{description}
\item[(C1)] $X$ and $Y$ are minimum;
\item[(C2)] $G$ contains a terrible cycle $C$ such that $X \cap V(C) \neq Y \cap V(C)$.
\end{description}

\begin{description}
\item[\textbf{Case 1.}] $G$ contains no terrible cycles:
For each path and cycle $C$ of $G$,
$X \cap V(C)$ and $Y \cap V(C)$ are $\max\{|X \cap V(C)|, |Y \cap V(C)|\}$-\TAR-reconfigurable by \cref{lem:path,lem:cycle}, respectively.
Concatenating such \TAR-sequences (where components $C$ with $|X \cap V(C)| \geq |Y \cap V(C)|$ are processed \emph{before} those with $|X \cap V(C)| < |Y \cap V(C)|$), we obtain a $k$-\TAR-sequence from $X$ to $Y$; i.e.,
$X$ and $Y$ are \TJ-reconfigurable due to \cref{obs:TJ-TAR}.

\item[\textbf{Case 2.}] $G$ contains terrible cycles, but $X$ and $Y$ are not minimum:
By assumption, we can find two elements $x \in X$ and $y \in Y$ such that 
$X' \triangleq X \setminus \{x\}$ and $Y' \triangleq Y \setminus \{y\}$ are target sets of size $k' \triangleq k-1$
in polynomial time by a brute-force search.
By \cref{lem:path,lem:cycle},
for each component $C$ of $G$,
$X' \cap V(C)$ and $Y' \cap V(C)$ are $(\max\{|X' \cap V(C)|, |Y' \cap V(C)|\}+1)$-\TAR-reconfigurable.
Concatenating such \TAR-sequences (in a similar manner to Case 1), we obtain
a $(k'+1)$-\TAR-sequence from $X'$ to $Y'$, implying that
$X$ and $Y$ are $k$-\TAR-reconfigurable; i.e.,
$X$ and $Y$ are \TJ-reconfigurable due to \cref{obs:TJ-TAR}.

\item[\textbf{Case 3.}] $G$ contains terrible cycles, and $X$ and $Y$ are minimum,
but it holds that $X \cap V(C) = Y \cap V(C)$ for every terrible cycle $C$:
Observe that there is no need to modify the vertices of terrible cycles.
For each path and nonterrible cycle $C$ of $G$,
$X \cap V(C)$ and $Y \cap V(C)$ are $\max\{|X \cap V(C)|, |Y \cap V(C)|\}$-\TAR-reconfigurable by \cref{lem:path,lem:cycle}, respectively.
Concatenating such \TAR-sequences (in a similar manner to Case 1), we obtain
a $k$-\TAR-sequence from $X$ to $Y$; i.e.,
$X$ and $Y$ are \TJ-reconfigurable due to \cref{obs:TJ-TAR}.

\item[\textbf{Case 4.}] Otherwise (i.e., (C1) and (C2) hold):
For a terrible cycle $C$ such that $X \cap V(C) \neq Y \cap V(C)$,
$X \cap V(C)$ and $Y \cap V(C)$ are not \TJ-reconfigurable on $C$ by \cref{lem:cycle}.
By \cref{lem:oplus},
$X$ and $Y$ are not \TJ-reconfigurable on $G$, as desired.
\end{description}
Since we can verify if $G$ satisfies (C1) and (C2) in polynomial time, the above analysis completes the proof.\footnote{
For example, \cref{fig:degree-2:X1,fig:degree-2:Y1} fall into Case 4;
\cref{fig:degree-2:X2,fig:degree-2:Y2} fall into Case 2.
}
\end{proof}

\subsection{\PSPACE-completeness on Planar $(3,3)$-Graphs \cite{hearn2005pspace,ito2011complexity,kaminski2012complexity}}
{Hearn and Demain}~\cite{hearn2005pspace} proved that \prb{Minimum Vertex Cover Reconfiguration}
is \PSPACE-complete on planar graphs of degree $2$ and $3$,
which implies that it is also \PSPACE-complete on planar $3$-regular graphs (see, e.g., \cite{mohar2001face}).
For the sake of completeness, we give an explicit proof of the following statement.
\begin{observation}[$\star$ \cite{hearn2005pspace,ito2011complexity,kaminski2012complexity}]
\label{thm:33}
\prb{Minimum Vertex Cover Reconfiguration} is \PSPACE-complete on planar $3$-regular graphs; i.e.,
\prb{Minimum Target Set Reconfiguration} is \PSPACE-complete on planar $(3,3)$-graphs.
\end{observation}

\subsection{\PSPACE-completeness on Bipartite Planar $(\{3,4\}, 2)$-Graphs}
We prove the \PSPACE-completeness result on bipartite planar $(\{3,4\}, 2)$-graphs.

\begin{theorem}
\label{thm:pb342}
\prb{Target Set Reconfiguration} is \PSPACE-complete on bipartite planar $(\{3,4\}, 2)$-graphs.
\end{theorem}

\begin{figure}[tbp]
\centering
\scalebox{0.7}{
    \begin{tikzpicture}
    \node[mynode] at (0,0) (vxy){$v_{xy}$};
    \node[mynode] at (-2,2) (vx){$v_x$};
    \node[mynode] at (-2,-2) (vy){$v_y$};
    \node[mynode] at (-4,2) (x){$x$};
    \node[mynode] at (-4,-2) (y){$y$};

    \node[mynode] at (5,0) (vw){$v_{w}$};
    \node[mynode] at (6.5,1.5) (w){$w$};
    \node[mynode] at (6.5,-1.5) (w1){$\overline{w}$};
    \node[mynode] at (8,0) (vz){$v_z$};
    \node[mynode] at (10,0) (z){$z$};

    \node[mynode] at (1,3) (hx){$h_x$};
    \node[mynode] at (2.5,4) (bx1){\Large $b_{x,1}$};
    \node[mynode] at (2.5,2) (bx2){\Large $b_{x,2}$};
    \node[mynode] at (4,3) (tx){$t_x$};

    \node[mynode] at (1,-3) (hy){$h_y$};
    \node[mynode] at (2.5,-2) (by1){\Large $b_{y,1}$};
    \node[mynode] at (2.5,-4) (by2){\Large $b_{y,2}$};
    \node[mynode] at (4,-3) (ty){$t_y$};
    
    \foreach \u/\v in {x/vx, y/vy, vx/vxy, vy/vxy, vxy/vw, vw/w, vw/w1, w/vz, w1/vz, vz/z, hx/bx1, hx/bx2, bx1/tx, bx2/tx, bx1/bx2, hy/by1, hy/by2, by1/ty, by2/ty, by1/by2}
        \draw[myedge] (\u)--(\v);
    \draw[myedge, rounded corners=4mm] (vx)|-(hx);
    \draw[myedge, rounded corners=4mm] (vy)|-(hy);
    \draw[myedge, rounded corners=4mm] (tx)-|(w);
    \draw[myedge, rounded corners=4mm] (ty)-|(w1);

    \draw[densely dashed, thick] \convexpath{hx,bx1,tx,bx2}{6mm};
    \draw[densely dashed, thick] \convexpath{hy,by1,ty,by2}{6mm};
    \draw[densely dashed, thick] \convexpath{w,vz,w1,vw}{6mm};
    \node[mylabel] at (4,1.5) {\LARGE $D_x$};
    \node[mylabel] at (4,-1.5) {\LARGE $D_y$};
    \node[mylabel] at (6.5,0) {\LARGE $C_4$};

    \node[mylabel, below=0.5mm of vx ]{$1$};
    \node[mylabel, above=0.5mm of vy ]{$1$};
    \node[mylabel, above=0.5mm of vxy]{$2$};
    \node[mylabel, right=0.5mm of vw]{$1$};
    \node[mylabel, below=0.5mm of w  ]{$2$};
    \node[mylabel, above=0.5mm of w1 ]{$2$};
    \node[mylabel, left =0.5mm of vz ]{$1$};
    \node[mylabel, right=0.5mm of hx ]{$2$};
    \node[mylabel, above=  2mm of bx1]{$1$};
    \node[mylabel, below=  2mm of bx2]{$1$};
    \node[mylabel, left =0.5mm of tx ]{$1$};
    \node[mylabel, right=0.5mm of hy ]{$2$};
    \node[mylabel, above=  2mm of by1]{$1$};
    \node[mylabel, below=  2mm of by2]{$1$};
    \node[mylabel, left =0.5mm of ty ]{$1$};
    \end{tikzpicture}
}
\caption{$\Upsilon$-gadget.}
\label{fig:upsilon}
\end{figure}
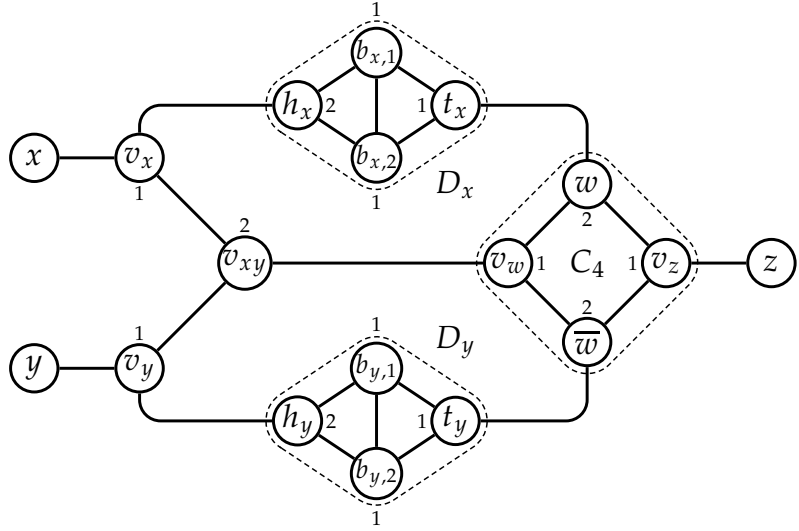

The proof of \cref{thm:pb342} is based on a series of reductions starting from a planar $(3,3)$-graph.
Suppose that $G=(V,E,\tau)$ is a planar $(3,3)$-graph and $w$ is a $(3,3)$-vertex, whose neighbors are denoted $N_G(w) = \{x,y,z\}$.
We then modify the subgraph induced by $\{w,x,y,z\}$ according to the following procedure (see \cref{fig:upsilon}).

\begin{shadebox}
\centering
\textbf{Construction of $\Upsilon$-gadget (\cref{fig:upsilon}).}
\begin{description}
\item[\textbf{Step 1.}] Remove $(x,w)$, $(y,w)$, and $(z,w)$, and set $\tau(w) = 2$.
\item[\textbf{Step 2.}] Create vertices $v_x, v_y, v_{xy}$ with
$\tau(v_x)=\tau(v_y)=1$ and $\tau(v_{xy})=2$, and
edges $(x,v_x)$, $(y,v_y)$, $(v_x, v_{xy})$, and $(v_y,v_{xy})$.
\item[\textbf{Step 3.}] Create a cycle graph $C_4$ on four vertices $w,v_z,\overline{w},v_w$ such that
$\tau(\overline{w})=2$ and $\tau(v_z)=\tau(v_w)=1$, and
edges $(v_w,v_{xy})$ and $(v_z,z)$.
\item[\textbf{Step 4.}]
Create a one-way gadget $D_x$ on vertex set $\{t_x,h_x,b_{x,1},b_{x,2}\}$ connecting from $w$ to $v_x$, and
create an edge $(b_{x,1},b_{x,2})$;
create a one-way gadget $D_y$ on vertex set $\{t_y,h_y,b_{y,1},b_{y,2}\}$ connecting from $\overline{w}$ to $v_y$ and
create an edge $(b_{y,1},b_{y,2})$.
\end{description}
\end{shadebox}
\noindent
We call the resulting subgraph a \emph{$\Upsilon$-gadget}, which plays a role in removing a $(3,3)$-vertex using $(3,1)$- and $(3,2)$-vertices without sacrificing planarity.
After uploading an early draft of this paper on arXiv,
Ryuhei Uehara discovered this gadget,
which is designed to preserve planarity, improving upon the old $\Upsilon$-gadget.
Gratefully, Uehara allowed us to include it here.
Vertices of $\{v_x,v_y,v_{xy}\} \uplus V(C_4) \uplus V(D_x) \uplus V(D_y)$ are referred to as \emph{internal vertices} of a $\Upsilon$-gadget.
Let $G'$ be a graph obtained from $G$ by the above procedure.
The most crucial property of $\Upsilon$-gadgets is that
$G'_{\{w\}} = G_{\{w\}}$.
For a seed set $S' \subseteq V'$ of $G'$,
we define $\phi_\Upsilon(S') \subseteq V$ as
\begin{align}
    \phi_\Upsilon(S') \triangleq
    \begin{cases}
    S' & \text{if } S' \subseteq V \setminus \{w\}, \\
    (S' \cap V) \cup \{w\} & \text{otherwise.}
    \end{cases}
\end{align}

\begin{lemma}
\label{lem:upsilon}
Let $G=(V,E,\tau)$ be a graph including a $(3,3)$-vertex $w$ and
$G'=(V',E',\tau')$ be a graph obtained from $G$ by replacing $w$ and its incident edges with a $\Upsilon$-gadget.
Then, a seed set $S' \subseteq V'$ is a minimum target set of $G'$
if and only if
$\phi_\Upsilon(S')$ is a minimum target set of $G$.
Moreover, two minimum target sets are \TJ-reconfigurable on $G$ if and only if
they are \TJ-reconfigurable on $G'$.
\end{lemma}
\begin{proof}
We give a relation between minimum target sets of $G$ and $G'$.
We first claim that if
a seed set $S \subseteq V$ is a target set of $G$,
then it is also a target set of $G'$
according to the following case analysis.
\begin{itemize}
    \item If $w \not\in S$:
    Since it holds that $d(w)=\tau(w)=3$,
    $S$ must be a target set of $G-\{w\}$.
    Observing that $G - \{w\}=G'[V \setminus \{w\}]$,
    we find $S$ to activate $x,y,z$ in $G'$,
    eventually activating the internal vertices of the $\Upsilon$-gadget:
    $x$ and $y$ respectively activate $v_x$ and $v_y$,
    which then activates $v_{xy}$ and $v_w$;
    $z$ activates $v_z$;
    $v_z$ and $v_w$ activate $w$ and $\overline{w}$, which further activates the vertices of $D_x$ and $D_y$.
    Therefore, $S$ is a target set of $G'$.

    \item If $w \in S$:
    By applying \cref{lem:residual} on the residual $G_{\{w\}}$ and $S$,
    we find $(S \setminus \{w\}) \cap V(G_{\{w\}}) = S \setminus \{w\}$ to be a target set of $G_{\{w\}}$.
    Since $G_{\{w\}} = G'_{\{w\}}$, $(S \setminus \{w\}) \uplus \{w\} = S$ is a target set of $G'$ by \cref{lem:residual}.
\end{itemize}

We then claim that if
$S' \subseteq V'$ is a target set of $G'$, then
$\phi_\Upsilon(S')$ is a target set of $G$.
To see this, the following case analysis is sufficient.

\begin{itemize}
    \item If $S' \subseteq V \setminus \{w\}$:
    Since $S'$ activates $x,y,z$ \emph{before} $w$ on $G'$,
    $S'$ is a target set of $G'[V \setminus \{w\}] = G-\{w\}$.
    Thus, $S'$ activates the vertices of $V \setminus \{w\}$ on $G$, finally activating $w$ on $G$;
    i.e., $S' = \phi_\Upsilon(S')$ is a target set of $G$.
    \item Otherwise:
    Observe that $(S' \cap V) \cup \{w\}$ is also a target set of $G'$.
    By applying \cref{lem:residual} on the residual $G'_{\{w\}}$ and $(S' \cap V) \cup \{w\}$,
    we find $S' \cap V$ to be a target set of $G'_{\{w\}}$.
    Since $G'_{\{w\}}=G_{\{w\}}$,
    $(S' \cap V) \cup \{w\} = \phi_\Upsilon(S')$ is a target set of $G$ by \cref{lem:residual}.
\end{itemize}
Note that a minimum target set $S' \subseteq V'$ of $G'$ includes
at most one internal vertex of the $\Upsilon$-gadget because otherwise
$(S' \cap V) \cup \{w\}$ is a target set of $G$,
which contradicts the minimality of $S'$.
Therefore, we have $|\phi_\Upsilon(S')| = |S'|$ for a minimum target set $S'$ of $G'$, namely, $\phi_\Upsilon(S')$ is a minimum target set of $G$.

We finally demonstrate that two minimum target sets $X$ and $Y$ of $G$ are \TJ-reconfigurable on $G$
if and only if they are \TJ-reconfigurable on $G'$.
Given a \TJ-sequence $\calS$ of minimum target sets of $G$ from $X$ to $Y$,
we find $\calS$ to be a \TJ-sequence of minimum target sets of $G'$ from $X$ to $Y$,
completing the only-if direction.
On the other hand, given a \TJ-sequence $\calS'$ of minimum target sets of $G'$ from $X$ to $Y$,
we find the sequence $\langle \phi_\Upsilon(S') \rangle_{S' \in \calS'}$
to be a \TJN-sequence of minimum target sets of $G$ from $\phi_\Upsilon(X)=X$ to $\phi_\Upsilon(Y)=Y$;
i.e., $X$ and $Y$ are \TJ-reconfigurable on $G$ by \cref{obs:tjn},
completing the if direction.
\end{proof}

\begin{figure}[tbp]
\centering
\scalebox{0.7}{
    \begin{tikzpicture}
    \input{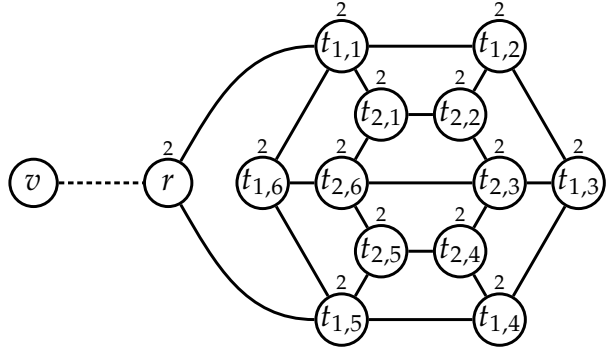}
    \node[mynode, left=16mm of r](v){$v$};
    \node[mylabel, fill=white, above=0.5mm of r]{$2$};
    \draw[myedge, densely dashed] (v)--(r);
    \end{tikzpicture}
}
\caption{$\Theta$-gadget.}
\label{fig:theta}
\end{figure}

After replacing each $(3,3)$-vertex and its incident edges with a $\Upsilon$-gadget,
we obtain a planar graph $H$ in which each vertex is
a $(2,1)$-, $(3,1)$-, $(3,2)$-, or $(4,2)$-vertex.
We then subdivide every edge according to \cref{lem:subdivision} to obtain
a bipartite planar graph $I$.
We further introduce the following gadget (see \cref{fig:theta}).

\begin{shadebox}
\centering
\textbf{Construction of $\Theta$-gadget (\cref{fig:theta}).}
\begin{description}
\item[\textbf{Step 1.}] Create a \emph{hexagonal prism graph} $Y_6$ on $12$ vertices such that
\begin{align}
    V(Y_6) & \triangleq \{t_{i,j} \mid i \in [2], j \in [6]\}, \\
    E(Y_6) & \triangleq \{ (t_{i,j}, t_{i,j+1\bmod 6}) \mid i \in [2], j \in [6] \} \uplus \{ (t_{1,j}, t_{2,j}) \mid j \in [6]\}, \\
    \tau(t_{i,j}) & \triangleq 2 \text{ for all } i \in [2], j \in [6].
\end{align}
\item[\textbf{Step 2.}] Create an edge $(t_{2,3}, t_{2,6})$.
\item[\textbf{Step 3.}] Create a vertex $r$ with $\tau(r)=2$ and two edges $(r, t_{1,1})$ and $(r, t_{1,5})$.
\end{description}
\end{shadebox}
\noindent
We call this gadget a \emph{$\Theta$-gadget}.
Observe that a $\Theta$-gadget is bipartite and planar.
We say that a $\Theta$-gadget is \emph{connected to} a vertex $v$ if
there exists an edge $(r,v)$.
We analyze the minimum target set of $\Theta$-gadgets, followed by \TJ-reconfigurability.

\begin{lemma}[$\star$]
\label{lem:mts-theta}
Let $R$ be a $\Theta$-gadget and
$R'$ be a graph obtained from $R$ by redefining the threshold of $r$ as $1$.
Then, the size of the minimum target set of $R$ and $R'$ is $3$.
In particular, the seed set $M \triangleq \{r,t_{1,2},t_{2,3}\}$ is a minimum target set of $R$ and $R'$.
\end{lemma}

\begin{lemma}
\label{lem:theta}
Let $G = (V,E,\tau)$ be a graph and 
$G' = (V',E',\tau')$ be a graph obtained from $G$ by
connecting a $\Theta$-gadget $R$ to a vertex $v$ of $G$ and
defining $\tau'(v) \triangleq \tau(v)+1$ and $\tau'(w) \triangleq \tau(w)$ for the other vertices $w \in V \setminus \{v\}$.
Then, a seed set $S \subseteq V$ is a minimum target set of $G$ if and only if
$S \uplus M$ is a minimum target set of $G'$, where
$M$ is the minimum target set of the $\Theta$-gadget given in \cref{lem:mts-theta}.
Moreover, two minimum target sets $X$ and $Y$ of $G$ are \TJ-reconfigurable on $G$
if and only if 
$X \uplus M$ and $Y \uplus M$ are \TJ-reconfigurable on $G'$.
\end{lemma}

\begin{remark}
The crux of the proof of \cref{lem:theta} is that
$R$ and $R'$ have the same-sized minimum target set as promised by \cref{lem:theta}.
Suppose we have constructed a different $\Theta$-gadget such that
the minimum target set of $R$ has size $3$ and that of $R'$ has size $2$.
Then, let $S'$ be any minimum target set of above defined $G'$, and
consider the residual $G'_{S'\cap V}$.
On one hand, if $v \in S'$,
the threshold of $r$ must be $1$ in $G'_{S'\cap V}$; i.e., $|S'\cap V(R)|=2$.
On the other hand, if $v \not\in S'$,
the threshold of $r$ must be $2$; i.e., $|S'\cap V(R)|=3$.
Therefore, $S'\cap V$ in the former case must not be a minimum target set of $G$.
This is undesirable.
\end{remark}

\begin{proof}[Proof of \cref{lem:theta}]
We give a relation between minimum target sets of $G$ and $G'$.
Let $R$ be a $\Theta$-gadget connected to $v$ and 
$R'$ be a graph obtained from $R$ by decreasing the threshold of $r$ by $1$.
Define $k$ as the size of the minimum target set of $G$.
Suppose that a seed set $S' \subseteq V'$ is a target set of $G'$.
Then, $S'$ satisfies the following:
\begin{description}
\item[\textbf{(C1)}] $S' \cap V$ is a target set of $G$.
In particular, it holds that $|S' \cap V| \geq k$.
\item[\textbf{(C2)}] $S' \cap V(R) $ is a target set of $R'$.
In particular, it holds that $|S' \cap V(R)| \geq 3$ due to \cref{lem:mts-theta}.
\end{description}
The reason (C1) holds is that
the residual $G'_{V(R)}$ is identical to $G$;
the reason (C2) holds is that
the residual $G'_{V}$ is identical to $R'$.
Indeed, for \emph{any} minimum target set $S \subseteq V$ of $G$ and
\emph{the} minimum target set $M$ of the $\Theta$-gadget given in \cref{lem:mts-theta},
the union $S \uplus M$ is a size-$(k+3)$ target set of $G'$ from the fact that $G'_M=G$. By (C1) and (C2), $S \uplus M$ turns out to be minimum.
Consequently, 
whenever $S'$ is a minimum target set of $G'$,
we have that
$S' \cap V$ is a minimum target set of $G$ due to (C1) and
$|S' \cap V(R)|=3$ due to (C2).

We finally demonstrate that two minimum target sets $X$ and $Y$ of $G$ are \TJ-reconfigurable
if and only if $X \uplus M$ and $Y \uplus M$ are \TJ-reconfigurable on $G'$.
Given a \TJ-sequence $\calS$ of minimum target sets of $G$ from $X$ to $Y$,
we find the sequence $\langle S \uplus M \rangle_{S \in \calS}$
to be a \TJ-sequence of minimum target sets of $G'$ from $X \uplus M$ to $Y \uplus M$ thanks to the above discussion.
On the other hand, given a \TJ-sequence $\calS'$ of minimum target sets of $G'$ from $X \uplus M$ to $Y \uplus M$,
we remove the vertices of $V(R)$ from every set in $\calS'$ to obtain
a new sequence $\calS = \langle S' \cap V \rangle_{S' \in \calS'}$.
Since each seed set in $\calS'$ is a union of a minimum target set of $G$ and a minimum target set of $R'$,
$\calS$ is a \TJN-sequence of minimum target sets of $G$ from $X$ to $Y$; i.e.,
$X$ and $Y$ are \TJ-reconfigurable on $G$ by \cref{obs:tjn}.
\end{proof}

We are now ready to prove \cref{thm:pb342} using \cref{lem:subdivision,lem:upsilon,lem:theta}.

\begin{proof}[Proof of \cref{thm:pb342}]
The reduction from a planar $(3,3)$-graph to a bipartite planar $(\{3,4\},2)$-graph is presented below.
\begin{shadebox}
\centering
\textbf{Reduction from planar $(3,3)$-graph $G$ to bipartite planar $(\{3,4\},2)$-graph $J$.}
\begin{description}
\item[\textbf{Step 1.}] Replace each vertex $w \in V(G)$ and its incident edges with a $\Upsilon$-gadget to obtain a graph $H$.
Note that $H$ is planar and that each vertex of $H$ is a
$(2,1)$-, $(3,1)$-, $(3,2)$-, or $(4,2)$-vertex.
\item[\textbf{Step 2.}] Subdivide each edge of $H$ by a new threshold-$1$ vertex to obtain a graph $I$.
Note that $I$ is bipartite and planar.
\item[\textbf{Step 3.}] Create a $\Theta$-gadget connecting to each $(2,1)$- or $(3,1)$- vertex $v$ of $I$, and
increase the threshold of $v$ by $1$ to obtain a graph $J$.
Note that $J$ is a bipartite planar $(\{3,4\}, 2)$-graph.
Let $M_v$ denote a minimum target set of the $\Theta$-gadget connected to vertex $v$
defined in \cref{lem:mts-theta}.
\end{description}
\end{shadebox}
\noindent
Obviously, the reduction completes in polynomial time, and
$J$ is a bipartite planar $(\{3,4\}, 2)$-graph.
We now show the correctness of the reduction.
Let $X$ and $Y$ be two minimum target sets of $G$.
By applying \cref{lem:subdivision,lem:upsilon} repeatedly, we have that
$X$ and $Y$ are \TJ-reconfigurable on $G$ if and only if
$X$ and $Y$ are \TJ-reconfigurable on $I$.
By applying \cref{lem:theta} repeatedly, we have that
$X$ and $Y$ are \TJ-reconfigurable on $I$ if and only if
$X_J \triangleq X \cup \bigcup_{v} M_v$ and
$Y_J \triangleq Y \cup \bigcup_{v} M_v$ are \TJ-reconfigurable on $J$.
Consequently, it turns out that
$X$ and $Y$ are \TJ-reconfigurable on $G$ if and only if
$X_J$ and $Y_J$ are \TJ-reconfigurable on $J$.
By \cref{thm:33}, \prb{Target Set Reconfiguration} on bipartite planar $(\{3,4\},2)$-graphs is \PSPACE-hard, as desired.
\end{proof}

\subsection{\PSPACE-completeness on Bipartite $(3,\{1,2\})$-Graphs and Planar $(3, \{1,2\})$-Graphs}
We next prove the \PSPACE-completeness result on
bipartite $(3, \{1,2\})$-graphs and planar $(3, \{1,2\})$-graphs.
\begin{theorem}
\label{thm:b312}
\prb{Target Set Reconfiguration} is \PSPACE-complete on
bipartite $(3,\{1,2\})$-graphs and planar $(3, \{1,2\})$-graphs.
\end{theorem}

Though \cref{thm:b312} is a more-or-less similar statement to \cref{thm:pb342}, its proof involves a different gadget.
Our reduction begins from a planar $(3,3)$-graph.
Suppose that $G=(V,E,\tau)$ is a planar $(3,3)$-graph and $w$ is a $(3,3)$-vertex, whose neighbors are denoted $N_G(w) = \{x,y,z\}$.
We then modify the subgraph induced by $\{w,x,y,z\}$
by replacing it with a $\Upsilon$-gadget (see \cref{fig:upsilon}).
We obtain a planar graph $H$ that consists only of $(3,1)$- or $(3,2)$-vertices.
We then make $H$ bipartite by subdividing every edge according to \cref{lem:subdivision}, which, however, produces $(2,1)$-vertices.
We thus introduce the following gadget (see \cref{fig:xi}):

\begin{figure}[tbp]
\centering
\scalebox{0.7}{
    \begin{tikzpicture}
    \node[mynode] at (-4,0) (a1){$a_1$};
    \node[mynode] at (-2.5,+1.5) (b1){$b_1$};
    \node[mynode] at (-1,0) (c1){$c_1$};
    \node[mynode] at (-2.5,-1.5) (d1){$d_1$};
    \node[mynode] at (+4,0) (a2){$a_2$};
    \node[mynode] at (+2.5,+1.5) (b2){$b_2$};
    \node[mynode] at (+1,0) (c2){$c_2$};
    \node[mynode] at (+2.5,-1.5) (d2){$d_2$};
    \node[mynode] at (-6,0) (v1){$v_1$};
    \node[mynode] at (+6,0) (v2){$v_2$};
    
    \node[mylabel, above=0.5mm of a1]{$2$};
    \node[mylabel, above=0.5mm of b1]{$1$};
    \node[mylabel, above=0.5mm of c1]{$1$};
    \node[mylabel, above=0.5mm of d1]{$1$};
    \node[mylabel, above=0.5mm of a2]{$2$};
    \node[mylabel, above=0.5mm of b2]{$1$};
    \node[mylabel, above=0.5mm of c2]{$1$};
    \node[mylabel, above=0.5mm of d2]{$1$};
    
    \foreach \u/\v in {a1/b1, b1/c1, c1/d1, d1/a1, a2/b2, b2/c2, c2/d2, d2/a2, b1/b2, c1/c2, d1/d2}
        \draw[myedge] (\u)--(\v);
    \draw[myedge, densely dashed] (v1)--(a1);
    \draw[myedge, densely dashed] (v2)--(a2);

    \draw[densely dashed, thick] \convexpath{a1,b1,c1,d1}{7.5mm};
    \draw[densely dashed, thick] \convexpath{a2,d2,c2,b2}{7.5mm};

    \node[mylabel] at (-2.5,0) {\LARGE $C_1$};
    \node[mylabel] at (+2.5,0) {\LARGE $C_2$};
    
    \end{tikzpicture}
}
\caption{$\Xi$-gadget.}
\label{fig:xi}
\end{figure}
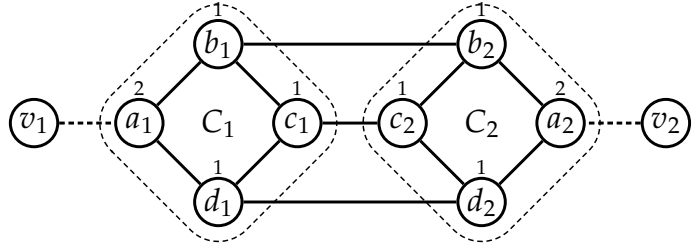

\begin{shadebox}
\centering
\textbf{Construction of $\Xi$-gadget (\cref{fig:xi}).}
\begin{description}
\item[\textbf{Step 1.}] Create two cycle graphs $C_1$ and $C_2$ on 
vertex sets $\{a_1,b_1,c_1,d_1\}$ and $\{a_2,b_2,c_2,d_2\}$, respectively, such that
$a_1$ and $a_2$ have a threshold of $2$ and the other vertices have a threshold of $1$.
\item[\textbf{Step 2.}] Create three edges $(b_1,b_2)$, $(c_1,c_2)$, and $(d_1,d_2)$.
\end{description}
\end{shadebox}
\noindent
We call this gadget a \emph{$\Xi$-gadget}.
Observe that a $\Xi$-gadget is bipartite.
We say that a $\Xi$-gadget \emph{connects between} two distinct vertices $v_1$ and $v_2$
if there exist two edges $(v_1, a_1)$ and $(v_2, a_2)$.
Vertices of $V(C_1) \uplus V(C_2)$ are referred to as \emph{internal vertices} of a $\Xi$-gadget.
We show the following lemma on \TJ-reconfigurability.

\begin{lemma}
\label{lem:xi}
Let $G=(V,E,\tau)$ be a graph and
$G'=(V',E',\tau')$ be a graph obtained from $G$ by
connecting a $\Xi$-gadget $R$ between two distinct vertices $v_1$ and $v_2$ of $G$ and
defining
$\tau'(v_1) \triangleq \tau(v_1) + 1$,
$\tau'(v_2) \triangleq \tau(v_2) + 1$, and 
$\tau'(w) \triangleq \tau(w)$ for the other vertices $w \in V \setminus \{v_1, v_2\}$.
Then, $S' \subseteq V'$ is a minimum target set of $G'$
if and only if
$S' \cap V$ is a minimum target set of $G$ and
$S' \cap V(R)$ is a minimum target set of $R$ consisting of a single internal vertex.
Moreover, two minimum target sets $X$ and $Y$ of $G$ are \TJ-reconfigurable on $G$ if and only if 
$X \uplus \{a_1\}$ and $Y \uplus \{a_1\}$ are \TJ-reconfigurable on $G'$.
\end{lemma}
\begin{proof}
We give a relation between minimum target sets of $G$ and $G'$.
Let $S' \subseteq V'$ be a minimum target set of $G'$.
Since $V$ is not a target set of $G'$, $S'$ includes at least one internal vertex of the $\Xi$-gadget.
Further, if $S'$ includes two or more internal vertices,
we can remove all but one of them to obtain a smaller target set.
Therefore, $S'$ must include \emph{exactly one} internal vertex, say, $v$.
Observing that $G'_{\{v\}} = G$ for any internal vertex $v$ of the $\Xi$-gadget,
we apply \cref{lem:residual} on the residual $G'_{\{v\}}$ and $S'$,
and find $S' \cap V$ to be a minimum target set of $G$.
On the other hand, if $S \subseteq V$ is a minimum target set of $G$,
then $S \uplus \{a_1\}$ is a minimum target set of $G'$.

We finally demonstrate that two minimum target sets $X$ and $Y$ of $G$ are \TJ-reconfigurable if and only if
$X \uplus \{a_1\}$ and $Y \uplus \{a_1\}$ are \TJ-reconfigurable on $G'$.
Given a \TJ-sequence $\calS$ of minimum target sets of $G$ from $X$ to $Y$,
we find the sequence $\langle S \uplus \{a_1\} \rangle_{S \in \calS}$ to be
a \TJ-sequence of minimum target sets of $G'$ from $X \uplus \{a_1\}$ to $Y \uplus \{a_1\}$,
completing the only-if direction.
On the other hand, given a \TJ-sequence $\calS'$ of minimum target sets of $G'$ from $X \uplus \{a_1\}$ to $Y \uplus \{a_1\}$,
we find the sequence $\langle S \cap V \rangle_{S \in \calS}$
to be a \TJN-sequence of minimum target sets of $G$ from $X$ to $Y$, i.e.,
$X$ and $Y$ are \TJ-reconfigurable on $G$, completing the if direction.
\end{proof}

We are now ready to prove \cref{thm:b312} using \cref{lem:upsilon,lem:xi,lem:subdivision,lem:oplus}.

\begin{proof}[Proof of \cref{thm:b312}]
The proof for planar $(3,\{1,2\})$-graphs is immediate from \cref{lem:upsilon}.
We present the reduction from a $(3,3)$-graph to a bipartite $(3,\{1,2\})$-graph below.
\begin{shadebox}
\centering
\textbf{Reduction from $(3,3)$-graph $G$ to bipartite $(3,\{1,2\})$-graph $J$.}
\begin{description}
\item[\textbf{Step 1.}] For each vertex $w$ of $G$, replace $w$ and its incident edges with a $\Upsilon$-gadget
to obtain a graph $H$. Note that $H$ is a $(3,\{1,2\})$-graph.
\item[\textbf{Step 2.}] Subdivide each edge of $H$ by a new threshold-$1$ vertex to obtain a graph $I$. Note that $I$ is bipartite.
\item[\textbf{Step 3.}] Create two copies of $I$, denoted $I_1$ and $I_2$.
For each vertex $v$ of $I$, the two vertices of $I_1$ and $I_2$ corresponding to $v$ are denoted by $v_1$ and $v_2$, respectively.
\item[\textbf{Step 4.}] For each $(2,1)$-vertex $v$ of $I$, connect a $\Xi$-gadget between $v_1$ and $v_2$, and increase the threshold of $v_1$ and $v_2$ by $1$ to obtain a graph $J$.
Denote by $a_{1,v}$ the internal vertex $a_1$ of the $\Xi$-gadget connected between $v_1$ and $v_2$.
\end{description}
\end{shadebox}
\noindent
Obviously, the reduction finishes in polynomial time, and
$J$ is a bipartite $(3,\{1,2\})$-graph.
Let $X$ and $Y$ be two minimum target sets of $G$.
By applying \cref{lem:upsilon,lem:subdivision,lem:oplus} repeatedly, we have that
$X$ and $Y$ are \TJ-reconfigurable on $G$ if and only if
$X_1 \uplus X_2$ and $Y_1 \uplus Y_2$ are \TJ-reconfigurable on $I_1 \oplus I_2$, where
$X_1 \triangleq \{v_1 \mid v \in X\}$,
$Y_1 \triangleq \{v_1 \mid v \in Y\}$,
$X_2 \triangleq \{v_2 \mid v \in X\}$, and
$Y_2 \triangleq \{v_2 \mid v \in Y\}$.
By applying \cref{lem:xi} repeatedly, we have that
$X_1 \uplus X_2$ and $Y_1 \uplus Y_2$ are \TJ-reconfigurable on $I_1 \oplus I_2$ if and only if
$X_J \triangleq X_1 \uplus X_2 \uplus \{a_{1,v} \mid v \in V(I)\}$ and
$Y_J \triangleq Y_1 \uplus Y_2 \uplus \{a_{1,v} \mid v \in V(I)\}$ are \TJ-reconfigurable on $J$.
Consequently,
$X$ and $Y$ are \TJ-reconfigurable on $G$ if and only if 
$X_J$ and $Y_J$ are \TJ-reconfigurable on $J$.
By \cref{thm:33},
\prb{Target Set Reconfiguration} on bipartite $(3,\{1,2\})$-graphs is \PSPACE-hard, as desired.
\end{proof}

\section{Restricted Graph Classes}
\label{sec:class}
This section investigates the tractability of
\prb{Target Set Reconfiguration} on restricted graph classes:
trees and split graphs.

\begin{algorithm}[tbp]
\caption{Chen's polynomial-time algorithm~\cite{chen2009approximability} for \prb{Target Set Selection} on a tree.}
\label{alg:chen-tree}
\begin{algorithmic}[1]
\Require tree $G=(V,E,\tau)$.
\State let $T$ be a rooted-tree representation of $G$ with root $r \in V$; initialize $S^* \leftarrow \emptyset$.
\While{$\exists$ vertex $v$ s.t.~$\tau'(v)$ is not defined, but $\tau'(w)$ is determined for all $v$'s children $w$}
    \State let $\tau'(v) \leftarrow \tau(v) - $
    (\# $v$'s children $w$ such that $\tau'(w) = 0$ or $w \in S^*$).
    \State \textbf{if} $v \neq r \textbf{ and } \tau'(v) \geq 2$ \textbf{then} $S^* \leftarrow S^* \cup \{v\}$. \label{linum:chen-tree:if1}
    \State \textbf{if} $v = r \textbf{ and } \tau'(v) \geq 1$ \textbf{then} $S^* \leftarrow S^* \cup \{v\}$.
\EndWhile
\State \textbf{return} $S^*$.
\end{algorithmic}
\end{algorithm}

\subsection{Polynomial Time on Trees}
\prb{Vertex Cover Reconfiguration} is known to be solvable in polynomial time on trees \cite{kaminski2012complexity,ito2016reconfiguration,mouawad2018vertex}.
We show that \prb{Target Set Reconfiguration} is also tractable on trees.

\begin{theorem}
\label{thm:tree}
\prb{Target Set Reconfiguration} is polynomial-time solvable on trees.
\end{theorem}

As will be shown, any pair of same-size target sets of a tree is \TJ-reconfigurable;
i.e., we just answer ``yes.''
Our idea for proving \cref{thm:tree} is
to construct a ``canonical'' target set that is \TAR-reconfigurable to any target set,
which is reminiscent of the idea for \prb{Dominating Set Reconfiguration} by
{Haddadan, Ito, Mouawad, Nishimura, Ono, Suzuki, and Tebbal}~\cite{haddadan2016complexity}.

\begin{figure}[tbp]
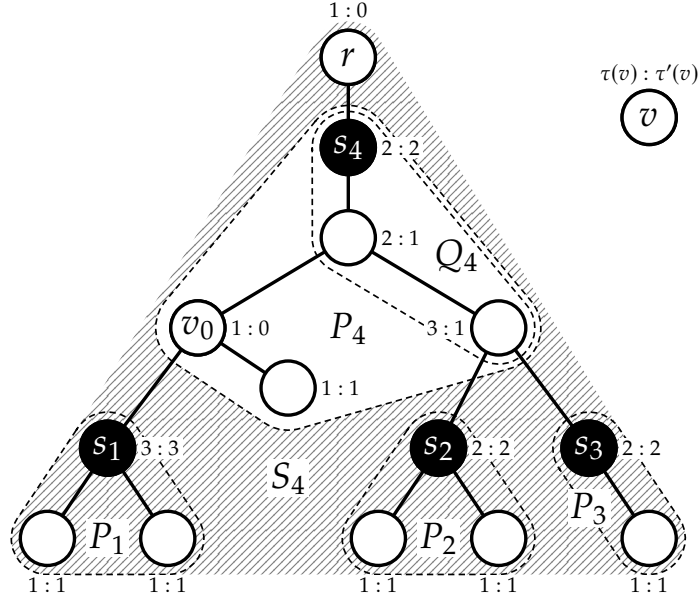

    \centering
    \scalebox{0.8}{\begin{tikzpicture}
        \input{fig_tree-temp1}
        \path[pattern=north east lines, pattern color=gray] \convexpath{r,h,m,i,e}{6mm};
        \draw[draw=none, fill=white] \convexpath{a,d,f,c}{7mm};
        \draw[densely dashed, thick] \convexpath{e,j,i}{6mm};
        \draw[densely dashed, thick] \convexpath{g,l,k}{6mm};
        \draw[densely dashed, thick] \convexpath{h,m}{6mm};
        \draw[densely dashed, thick] \convexpath{a,d,f,c}{7mm};
        \draw[densely dashed, thick] \convexpath{a,d,b}{6mm};
        \input{fig_tree-temp2}
        \node[inner sep=0.5mm, fill=white] at (-4,-6.5) {\LARGE $P_1$};
        \node[inner sep=0.5mm, fill=white] at (1.5,-6.5) {\LARGE $P_2$};
        \node[inner sep=0.5mm, fill=white] at (4,-6) {\LARGE $P_3$};
        \node[inner sep=0.5mm, fill=white] at (0,-3) {\LARGE $P_4$};
        \node[inner sep=0.5mm] at (1.8,-1.8) {\LARGE $Q_4$};
        \node[inner sep=0.5mm, fill=white] at (-1,-5.5) {\LARGE $S_4$};
    \end{tikzpicture}}
    \caption{
        Chen's algorithm \cite{chen2009approximability} on a tree $G$ rooted at $r$, and
        the proof of \cref{lem:tree-chara}.
        Each vertex $v$ is associated with a pair of $\tau(v)$ and $\tau'(v)$,
        denoted $\tau(v) : \tau'(v)$.
        In a bottom-up fashion,
        we compute $\tau'(v)$ as $\tau(v)$ minus the number of $v$'s children
        that would have been activated by the intermediate solution.
        Selected vertices $S^* = \{s_1,s_2,s_3,s_4\}$ are colored black.
        Each dashed line represents $P_1,P_2,P_3,P_4,$ or $Q_4$;
        $S_4 = V(G) \setminus P_4$ is denoted by diagonal lines;
        $S'_4$ is defined as $S_4 \uplus \{v_0\}$, which turns out not to be a target set.
    }
    \label{fig:tree}
\end{figure}

We first recapitulate Chen's polynomial-time algorithm~\cite{chen2009approximability} for \prb{Target Set Selection} on a tree, presented in \cref{alg:chen-tree}.
Let $G=(V,E,\tau)$ be a tree.
For an arbitrary vertex $r \in V$,
let $T$ denote a tree representation of $(V,E)$ rooted at $r$,
which naturally introduces parents, children, and leaves.
Starting from an empty set $S^* = \emptyset$,
we determine whether or not to include each vertex of $G$ into $S^*$ in a bottom-up fashion.
Suppose that there exists a vertex $v$ that has not been scanned yet but its children have already been examined.\footnote{
Note that we first select a leaf, which has no children.}
We then define $\tau'(v)$ as follows:
\begin{align}
\tau'(v) \triangleq \tau(v) - \left|\Bigl\{ \text{child } w \text{ of } v \mid \tau'(w)=0 \text{ or } w \in S^* \Bigr\}\right|,
\end{align}
which indicates
the number of $v$'s children that would have been activated by the current $S^*$.
If $v$ is not the root $r$, then
we include $v$ into $S^*$ only if $\tau'(v) \geq 2$.
On the other hand, if $v = r$, then
we include $v$ into $S^*$ only if $\tau'(v) \geq 1$.
Running through every vertex of $G$, we finally return $S^*$ as an output.
See \cref{fig:tree} for a running example of \cref{alg:chen-tree}.
Chen \cite{chen2009approximability} proves that $S^*$ is a minimum target set of $G$.
We will present a simple characterization of target sets of $G$ using $S^*$.
Define $k \triangleq |S^*|$.
Let $s_1, s_2, \ldots, s_k$ denote the vertices of $S^*$ ordered by
a postorder depth-first traversal starting from the root $r$ of $T$.
For each $i \in [k]$,
let $T_i$ denote the subtree of $T$ rooted at $s_i$, and
define the vertex set $P_i$ as:
\begin{align}
    P_i \triangleq V(T_i) \setminus \bigcup_{j \in [i-1]} V(T_j).
\end{align}
$\{P_1, \ldots, P_k\}$ forms a \emph{packing} of $V(T)$, and
it holds that $S^* \cap P_i = \{s_i\}$ for all $i \in [k]$; see also \cref{fig:tree}.

\begin{lemma}
\label{lem:tree-chara}
For any target set $S$ of a tree $G$,
it holds that $|S \cap P_i| \geq 1$ for all $i \in [k]$.
\end{lemma}
\begin{proof}
We show that for each $i \in [k]$,
the seed set $S_i \triangleq V(G) \setminus P_i$ is not a target set, which is sufficient because
$S_i$ is the maximum set disjoint to $P_i$.
In fact, we show that $S'_i \triangleq S_i \uplus \{v \in P_i \mid \tau'(v) = 0\}$ is not a target set,
where the values of $\tau'$ are obtained by running \cref{alg:chen-tree}.
Fix $i$ and consider first the case of $s_i \neq r$.
By definition of $\tau'$ and $S'_i$,
we have the following relation for each $v \in P_i$:
\begin{itemize}
    \item If $\tau'(v) = 0$: $v \in S'_i$;
    \item If $\tau'(v) = 1$: $\tau(v)$ is equal to the number of $v$'s children in $S'_i$ plus $1$;
    \item If $\tau'(v) \geq 2$: $\tau(v)$ is at least the number of $v$'s children in $S'_i$ plus $2$.
\end{itemize}
Define $Q_i$ as the set of vertices that are reachable from $s_i$ in $T$ without touching any vertices of $S'_i$.
Observe that $s_i \in Q_i \subseteq P_i$, and the subtree of $T$ induced by $Q_i$ is connected.
See \cref{fig:tree} for an example.

For each vertex $v \in Q_i \setminus \{s_i\}$,
$v$'s parent is \emph{not} in $S'_i$, and
$\tau'(v)$ must be $1$ because
in Line~\ref{linum:chen-tree:if1} of \cref{alg:chen-tree} we have not added $v$ into $S^*$;
thus, we have $\tau(v) = |N_G(v) \cap S'_i| + 1$.
Similarly,
$s_i$'s parent is in $S'_i$, and
$\tau'(s_i)$ must be at least $2$ because 
in Line~\ref{linum:chen-tree:if1} of \cref{alg:chen-tree} we have added $s_i$ into $S^*$;
hence, $\tau(s_i) \geq |N_G(s_i) \cap S'_i| + 1$.

Consequently, $\tau(v) \geq |N_G(v) \cap S'_i| + 1$ for all $v \in Q_i$, implying that
no vertices of $Q_i$ would be activated; i.e.,
$S'_i$ is not a target set, as desired.
The case of $s_i=r$ can be shown in the same manner
except that $\tau'(r)$ is at least the number of $r$'s children in $S'_i$ plus $1$, which eventually results in that $\tau(r) \geq |N_G(r) \cap S'_i| + 1$ because $r$ has no parent.
\end{proof}

\cref{lem:tree-chara} gives a different (perhaps simple) proof of the optimality of \cref{alg:chen-tree} from Chen's proof.
Indeed, we claim that any target set $S$ and $S^*$ are $|S|$-\TAR-reconfigurable.

\begin{lemma}
\label{lem:tree-tar}
For any target set $S$ of $G$,
$S$ and $S^*$ are $|S|$-\TAR-reconfigurable.
Moreover, an actual $|S|$-\TAR-sequence can be found in polynomial time.
\end{lemma}
\begin{proof}
We construct a \TAR-sequence $\calS$ of target sets from a target set $S$ to $S^*$.
Starting from $S^{(0)} \triangleq S$, for each $i \in [k]$,
we transform $S^{(i-1)}$ into $S^{(i)}$ by the following \TAR-steps:
\begin{shadebox}
\begin{description}
\item[\textbf{Step 1.}] Add the vertex $s_i \in S^* \cap P_i$ if $s_i \not\in S^{(i-1)}$.
\item[\textbf{Step 2.}] Remove the vertices of $S^{(i-1)} \cap (P_i \setminus \{s_i\})$ one by one.
\item[\textbf{Step 3.}] Let $S^{(i)}$ be the resulting set. Note that $S^{(i)} \cap V(T_i) = S^* \cap V(T_i)$.
\end{description}
\end{shadebox}
\noindent Since it holds that $S^* \subseteq S^{(k)}$,
we finally remove the vertices of $S^{(k)} \setminus \bigcup_{i \in [k]} P_i$ to obtain $S^{(k+1)} = S^*$.

We here show that $S^{(i)}$ is a target set for each $i \in [k]$.
The proof is done by induction on $i$.
The base case of $i=0$ is obvious since $S^{(0)}=S$.
Suppose that $S^{(i-1)}$ is a target set for $i \in [k]$.
Since $G$ is a tree, the residual $G_{\{s_i\}}$ can be decomposed into
the subtree $G_1$ of $G_{\{s_i\}}$ induced by $V(T_i) \setminus \{s_i\}$ and
the subtree $G_2$ of $G_{\{s_i\}}$ induced by $V(G) \setminus V(T_i)$.
Note that $S^{(i)} \cap V(G_1) = S^* \cap V(G_1)$ and 
$S^{(i)} \cap V(G_2) = S^{(i-1)} \cap V(G_2)$.
Since $S^*$ is a target set of $G$,
$S^* \setminus \{s_i\}$ is a target set of $G_{\{s_i\}}$ by \cref{lem:residual}, which implies that
$(S^* \setminus \{s_i\}) \cap V(G_1) = S^* \cap V(G_1) = S^{(i)} \cap V(G_1)$ is a target set of $G_1$.
By the induction hypothesis, $S^{(i-1)} \cup \{s_i\}$ is a target set of $G$,
implying that $(S^{(i-1)} \cup \{s_i\}) \cap V(G_2)$ is a target set of $G_2$ by \cref{lem:residual}; i.e.,
$(S^{(i-1)} \cup s_i) \cap V(G_2) = S^{(i-1)} \cap V(G_2) = S^{(i)} \cap V(G_2)$ is a target set of $G_2$.
Eventually, we have that
$S^{(i)} \cap (V(G_1) \cup V(G_2)) = S^{(i)} \setminus \{s_i\}$ is a target set of
$G_1 \oplus G_2 = G_{\{s_i\}}$; i.e.,
$S^{(i)}$ is a target set of $G$ due to \cref{lem:residual}.
Obviously, $S^{(k+1)} = S^*$ is a target set.
Since every seed set appearing in $\calS$ is a superset of $S^{(i)}$ for some $i \in [k+1]$, $\calS$ is a \TAR-sequence of target sets from $S$ to $S^*$. 

We then claim that $|S^{(i-1)}| \geq |S^{(i)}|$ for all $i \in [k]$.
If $s_i \in S^{(i-1)}$, then the claim is obvious because we only remove vertices in Step 2 without adding $s_i$ in Step 1.
On the other hand, if $s_i \not\in S_{i-1}$,
\cref{lem:tree-chara} tells that $S^{(i-1)} \cap (P_i \setminus \{s_i\}) \neq \emptyset$.
Hence, we remove at least one vertex in Step 2; i.e.,
it must hold that $|S^{(i-1)}| \geq |S^{(i)}|$.
It is easy to observe that $|S^{(k)}| \geq |S^{(k+1)}|$.
Since every target set in the subsequence of $\calS$
from $S^{(i-1)}$ to $S^{(i)}$ has a size of at most $|S^{(i-1)}|+1$,
the maximum size of any target set in $\calS$ from
$S^{(0)} = S$ to $S^{(k+1)} = S^*$ is at most $|S|+1$;
i.e., $\calS$ is a $|S|$-\TAR-sequence from $S$ to $S^*$, completing the proof.
\end{proof}

\begin{proof}[Proof of \cref{thm:tree}]
By \cref{lem:tree-tar},
two target sets $X$ and $Y$ are $\max\{|X|, |Y|\}$-\TAR-reconfigurable.
In particular, when $|X|=|Y|$, they are \TJ-reconfigurable,
as desired.
\end{proof}

\subsection{\PSPACE-completeness on Split Graphs}
A graph is called a \emph{split graph} if the vertex set can be partitioned into a clique and an independent set.
On split graphs, \prb{VC-R} is solvable in polynomial time \cite{kaminski2012complexity,ito2016reconfiguration,mouawad2018vertex}.
But,
\prb{Target Set Reconfiguration} is \PSPACE-complete on split graphs.

\begin{theorem}
\label{thm:split}
\prb{Target Set Reconfiguration} is \PSPACE-complete on split graphs.
\end{theorem}

We adapt a reduction from \prb{Hitting Set} to \prb{Target Set Selection} due to
{Nichterlein, Niedermeier, Uhlmann, and Weller}~\cite{nichterlein2013tractable}.
Given a set family $\calF = \{F_1, \ldots, F_m\}$ over a universe $U = \{u_1, \ldots, u_n\}$,
a subset $S$ of $U$ is called a \emph{hitting set} if 
$S \cap F_j \neq \emptyset$ for all $j \in [m]$.
The \prb{Hitting Set} problem requires deciding if there exists a hitting set of size $k$ for a parameter $k \in [n]$, which is known to be \NP-complete \cite{johnson1979computers}.
In \prb{Hitting Set Reconfiguration}, 
given $\calF$, $U$, and two hitting sets $S$ and $T$ of size $k$,
we are requested to determine the existence of
a \TJ-sequence of hitting sets from $S$ to $T$.
Due to the equivalence between \prb{Set Cover} and \prb{Hitting Set},
\prb{Hitting Set Reconfiguration} can be shown to be \PSPACE-complete \cite{ito2011complexity}.
Given $U$, $\calF$, and a set size $k \in [n]$,
we construct a graph $G = (V,E,\tau)$ according to the following procedure \cite{nichterlein2013tractable}:

\begin{shadebox}
\centering
\textbf{Construction of $G=(V,E,\tau)$ from $U, \calF, k$ \cite{nichterlein2013tractable}.}
\begin{description}
\item[\textbf{Step 1.}] Create vertex sets
$V_U \triangleq \{v_u \mid u \in U\}$, $W_\calF \triangleq \{ w_F : F \in \calF \}$, and 
an isolated vertex $x$. Define $V \triangleq V_U \uplus W_\calF \uplus \{x\}$.
\item[\textbf{Step 2.}] Create an edge $(v_u, w_F)$ for each $v_u \in V_U$ and $w_F \in W_\calF$ such that $u \in F$.
\item[\textbf{Step 3.}] Connect $x$ to all vertices in $W_\calF$; i.e., create edges $(x, w_F)$ for each $w_F \in W_\calF$.
\item[\textbf{Step 4.}] Render $V_U \uplus \{x\}$ a clique; i.e., create edges between every pair of vertices of $V_U \uplus \{x\}$.
\item[\textbf{Step 5.}] Set
$\tau(v_u) \triangleq |\{F \in \calF \mid u \in F\}| + k + 1$ for each $v_u \in V_U$,
$\tau(w_F) \triangleq 1$ for each $w_F \in W_\calF$, and
$\tau(x) \triangleq |W_\calF|+k$.
\end{description}
\end{shadebox}

Since $V_U \uplus \{x\}$ forms a clique and $W_\calF$ forms an independent set, $G$ is a split graph.
Moreover, the diameter of $G$ is $2$ as $x$ is adjacent to every other vertex.
{Nichterlein, Niedermeier, Uhlmann, and Weller}~\cite{nichterlein2013tractable} proved that
there exists a size-$k$ hitting set if and only if
there exists a size-$k$ target set of $G$,
implying the \NP-hardness of \prb{Target Set Selection} on split graphs of diameter $2$.
In particular, we use the following fact to prove \cref{thm:split}.

\begin{lemma}[{Nichterlein et al.}~\cite{nichterlein2013tractable}]
\label{lem:split}
For a set family $\calF = \{F_1, \ldots, F_m\}$ of
a universe $U = \{u_1, \ldots, u_n\}$ and a positive integer $k \in [n]$,
let $G=(V,E,\tau)$ be a graph constructed from $U, \calF, k$ according to the procedure described above.
Then, any size-$k$ target set $S$ of $G$ is a subset of $V_U$; i.e., it does not intersect $W_F$ or include $x$.
Moreover, $S \subseteq U$ is a size-$k$ hitting set if and only if
$V_S \triangleq \{v_u \mid u \in S\}$ is a size-$k$ target set.
\end{lemma}
\begin{proof}
In \cite[Proof of Theorem 1]{nichterlein2013tractable},
it is shown that a size-$k$ target set $S$ never includes the vertex $x$.
Suppose then that $S$ includes a vertex $w_F$ of $W_\calF$.
By \cref{obs:replace} and the fact that $\tau(w_F) = 1$,
$S \setminus \{w\} \cup \{x\}$ must be a target set, which is a contradiction.
See \cite[Proof of Theorem 1]{nichterlein2013tractable} for
the proof of the equivalence between a size-$k$ hitting set and a size-$k$ target set.
\end{proof}

\begin{proof}[Proof of \cref{thm:split}]
We present a polynomial-time reduction from
\prb{Hitting Set Reconfiguration}, which is a \PSPACE-complete problem \cite{ito2011complexity}.
Let $\calF$ be a set family of a universe $U$ and
$X$ and $Y$ be two hitting sets of size $k$.
Let $G$ be a graph constructed from $U$, $\calF$, $k$ according to the procedure described above in polynomial time.
Define
$V_X \triangleq \{v_x \in V_U \mid x \in X\}$ and
$V_Y \triangleq \{v_y \in V_U \mid y \in Y\}$.
Given a \TJ-sequence $\calS$ of size-$k$ hitting sets from $X$ to $Y$,
we have that the sequence
$\langle \{ v_u \mid u \in S \} \rangle_{S \in \calS}$ is a \TJ-sequence of size-$k$ target sets from $V_X$ to $V_Y$ by \cref{lem:split}.
On the other hand,
given a \TJ-sequence $\calT$ from $V_X$ to $V_Y$,
we have that the sequence $\langle \{ u \mid v_u \in T \} \rangle_{T \in \calT}$
is a \TJ-sequence of size-$k$ hitting sets from $X$ to $Y$ by \cref{lem:split}.
Consequently, $X$ and $Y$ are \TJ-reconfigurable on $\calF$ if and only if
$V_X$ and $V_Y$ are \TJ-reconfigurable on $G$, which completes the proof.
\end{proof}

\section{Discussion}
Our results follow a typical pattern, that is,
that an \NP-complete (resp.~\cP) search problem induces a \PSPACE-complete (resp.~\cP) reconfiguration problem, which however left some open questions.
One of the important unsettled cases is cubic graphs of threshold $2$,
which is equivalent to \prb{FVS-R} on cubic graphs and
was mentioned by Suzuki in the open problem session of the 3rd International Workshop on Combinatorial Reconfiguration.\footnote{\url{https://pagesperso.g-scop.grenoble-inp.fr/~bousquen/CoRe_2019/CoRe_2019_Open_Problems.pdf}}
Since the respective search problem can be solved in polynomial time using a \emph{graphic matroid parity} algorithm \cite{takaoka2015note,kyncl2017irreversible},
a polynomial-time algorithm might be expected.
We stress that \PSPACE-completeness has been shown for cubic graphs of threshold $1$ and $2$ in this paper.
A superclass of trees is another case whose complexity remains open; e.g.,
\prb{VC-R} is known to be polynomial-time solvable on cacti \cite{mouawad2018vertex} but is \PSPACE-complete on $O(1)$-treewidth graphs~\cite{wrochn2018reconfiguration}.
The complexity status for claw-free graphs \cite{bonsma2014reconfiguring,munaro2017line} is also left unanswered.

\section*{Acknowledgements}
I thank
members of the project \emph{Fusion of Computer Science, Engineering and Mathematics Approaches for Expanding Combinatorial Reconfiguration}
for giving me an opportunity to talk about this paper at the 20\textsuperscript{th} CoRe Seminar in January 2022;
I especially want to thank
Takehiro Ito and Naonori Kakimura for inviting me to the seminar, and
Ryuhei Uehara for allowing me to share the $\Upsilon$-gadget of \cref{lem:upsilon},
which is an elegant improvement upon the old $\Upsilon$-gadget appearing in an early draft of this paper.
Also, I thank the anonymous referees for many suggestions which made the presentation of this paper much better.
This work was partially done while the author was at NEC.

\bibliographystyle{alpha}
\bibliography{manu}

\newcommand{\etalchar}[1]{$^{#1}$}
\begin{thebibliography}{MNRW14}

\bibitem[ABW10]{ackerman2010combinatorial}
Eyal Ackerman, Oren Ben{-}Zwi, and Guy Wolfovitz.
\newblock Combinatorial model and bounds for target set selection.
\newblock {\em Theor. Comput. Sci.}, 411(44-46):4017--4022, 2010.

\bibitem[BC09]{bonsma2009finding}
Paul Bonsma and Luis Cereceda.
\newblock Finding paths between graph colourings: {PSPACE}-completeness and
  superpolynomial distances.
\newblock {\em Theor. Comput. Sci.}, 410(50):5215--5226, 2009.

\bibitem[BCNS14a]{bazgan2014parameterized}
Cristina Bazgan, Morgan Chopin, Andr{\'e} Nichterlein, and Florian Sikora.
\newblock Parameterized approximability of maximizing the spread of influence
  in networks.
\newblock {\em J. Discrete Algorithms}, 27:54--65, 2014.

\bibitem[BCNS14b]{bazgan2014parameterizeda}
Cristina Bazgan, Morgan Chopin, Andr{\'{e}} Nichterlein, and Florian Sikora.
\newblock Parameterized inapproximability of target set selection and
  generalizations.
\newblock {\em Comput.}, 3(2):135--145, 2014.

\bibitem[BEPR19]{bessy2019dynamic}
St{\'e}phane Bessy, Stefan Ehard, Lucia~D. Penso, and Dieter Rautenbach.
\newblock Dynamic monopolies for interval graphs with bounded thresholds.
\newblock {\em Discrete Appl. Math.}, 260:256--261, 2019.

\bibitem[BHLN11]{ben-zwi2011treewidth}
Oren Ben{-}Zwi, Danny Hermelin, Daniel Lokshtanov, and Ilan Newman.
\newblock Treewidth governs the complexity of target set selection.
\newblock {\em Discrete Optim.}, 8(1):87--96, 2011.

\bibitem[BKW14]{bonsma2014reconfiguring}
Paul Bonsma, Marcin Kami{\'n}ski, and Marcin Wrochna.
\newblock Reconfiguring independent sets in claw-free graphs.
\newblock In {\em SWAT}, volume 8503, pages 86--97, 2014.

\bibitem[Bon13]{bonsma2013complexity}
Paul Bonsma.
\newblock The complexity of rerouting shortest paths.
\newblock {\em Theor. Comput. Sci.}, 510:1--12, 2013.

\bibitem[Bon16]{bonsma2016independent}
Paul Bonsma.
\newblock Independent set reconfiguration in cographs and their
  generalizations.
\newblock {\em J. Graph Theory}, 83(2):164--195, 2016.

\bibitem[CDP{\etalchar{+}}11]{centeno2011irreversible}
Carmen~C. Centeno, Mitre~Costa Dourado, Lucia~Draque Penso, Dieter Rautenbach,
  and Jayme~Luiz Szwarcfiter.
\newblock Irreversible conversion of graphs.
\newblock {\em Theor. Comput. Sci.}, 412(29):3693--3700, 2011.

\bibitem[CFK{\etalchar{+}}15]{cygan2015parameterized}
Marek Cygan, Fedor~V. Fomin, {\L}ukasz Kowalik, Daniel Lokshtanov, D{\'a}niel
  Marx, Marcin Pilipczuk, Micha{\l} Pilipczuk, and Saket Saurabh.
\newblock {\em Parameterized Algorithms}.
\newblock Springer, 2015.

\bibitem[Che09]{chen2009approximability}
Ning Chen.
\newblock On the approximability of influence in social networks.
\newblock {\em {SIAM} J. Discrete Math.}, 23(3):1400--1415, 2009.

\bibitem[CHL{\etalchar{+}}13]{chiang2013some}
Chun-Ying Chiang, Liang-Hao Huang, Bo-Jr Li, Jiaojiao Wu, and Hong-Gwa Yeh.
\newblock Some results on the target set selection problem.
\newblock {\em J. Comb. Optim.}, 25(4):702--715, 2013.

\bibitem[CNNW14]{chopin2014constant}
Morgan Chopin, Andr{\'e} Nichterlein, Rolf Niedermeier, and Mathias Weller.
\newblock Constant thresholds can make target set selection tractable.
\newblock {\em Theory Comput. Syst.}, 55(1):61--83, 2014.

\bibitem[CNW16]{charikar2016approximating}
Moses Charikar, Yonatan Naamad, and Anthony Wirth.
\newblock On approximating target set selection.
\newblock In {\em APPROX/RANDOM}, pages 4:1--4:16, 2016.

\bibitem[CvdHJ08]{cereceda2008connectedness}
Luis Cereceda, Jan van~den Heuvel, and Matthew Johnson.
\newblock Connectedness of the graph of vertex-colourings.
\newblock {\em Discrete Math.}, 308(5-6):913--919, 2008.

\bibitem[CvdHJ11]{cereceda2011finding}
Luis Cereceda, Jan van~den Heuvel, and Matthew Johnson.
\newblock Finding paths between 3-colorings.
\newblock {\em J. Graph Theory}, 67(1):69--82, 2011.

\bibitem[DDF{\etalchar{+}}15]{demaine2015linear}
Erik~D. Demaine, Martin~L. Demaine, Eli Fox{-}Epstein, Duc~A. Hoang, Takehiro
  Ito, Hirotaka Ono, Yota Otachi, Ryuhei Uehara, and Takeshi Yamada.
\newblock Linear-time algorithm for sliding tokens on trees.
\newblock {\em Theor. Comput. Sci.}, 600:132--142, 2015.

\bibitem[DF12]{downey2012parameterized}
Rodney~G. Downey and Michael~R. Fellows.
\newblock {\em Parameterized Complexity}.
\newblock Springer, 2012.

\bibitem[DKT18]{dvorak2018target}
Pavel Dvo\v{r}{\'a}k, Dusan Knop, and Tom{\'a}\v{s} Toufar.
\newblock Target set selection in dense graph classes.
\newblock In {\em ISAAC}, volume 123, pages 18:1--18:13, 2018.

\bibitem[DR09]{dreyer2009irreversible}
Paul~A. Dreyer, Jr. and Fred~S. Roberts.
\newblock Irreversible $k$-threshold processes: Graph-theoretical threshold
  models of the spread of disease and of opinion.
\newblock {\em Discrete Appl. Math.}, 157(7):1615--1627, 2009.

\bibitem[Dre00]{dreyer2000applications}
Paul~Andrew Dreyer, Jr.
\newblock {\em Applications and Variations of Domination in Graphs}.
\newblock PhD thesis, Rutgers University New Jersey, 2000.

\bibitem[FK19]{feige2019target}
Uriel Feige and Shimon Kogan.
\newblock Target set selection for conservative populations.
\newblock {\em CoRR}, abs/1909.03422, 2019.

\bibitem[GKMP09]{gopalan2009connectivity}
Parikshit Gopalan, Phokion~G. Kolaitis, Elitza Maneva, and Christos~H.
  Papadimitriou.
\newblock The connectivity of {Boolean} satisfiability: Computational and
  structural dichotomies.
\newblock {\em SIAM J. Comput.}, 38(6):2330--2355, 2009.

\bibitem[HD05]{hearn2005pspace}
Robert~A. Hearn and Erik~D. Demaine.
\newblock {PSPACE}-completeness of sliding-block puzzles and other problems
  through the nondeterministic constraint logic model of computation.
\newblock {\em Theor. Comput. Sci.}, 343(1-2):72--96, 2005.

\bibitem[HD09]{hearn2009games}
Robert~A. Hearn and Erik~D. Demaine.
\newblock {\em Games, Puzzles, and Computation}.
\newblock A K Peters, Ltd., 2009.

\bibitem[HIM{\etalchar{+}}16]{haddadan2016complexity}
Arash Haddadan, Takehiro Ito, Amer~E. Mouawad, Naomi Nishimura, Hirotaka Ono,
  Akira Suzuki, and Youcef Tebbal.
\newblock The complexity of dominating set reconfiguration.
\newblock {\em Theor. Comput. Sci.}, 651:37--49, 2016.

\bibitem[ID14]{ito2014approximability}
Takehiro Ito and Erik~D. Demaine.
\newblock Approximability of the subset sum reconfiguration problem.
\newblock {\em J. Comb. Optim.}, 28(3):639--654, 2014.

\bibitem[IDH{\etalchar{+}}11]{ito2011complexity}
Takehiro Ito, Erik~D. Demaine, Nicholas J.~A. Harvey, Christos~H.
  Papadimitriou, Martha Sideri, Ryuhei Uehara, and Yushi Uno.
\newblock On the complexity of reconfiguration problems.
\newblock {\em Theor. Comput. Sci.}, 412(12-14):1054--1065, 2011.

\bibitem[INZ16]{ito2016reconfiguration}
Takehiro Ito, Hiroyuki Nooka, and Xiao Zhou.
\newblock Reconfiguration of vertex covers in a graph.
\newblock {\em {IEICE} Trans. Inf. Syst.}, 99-D(3):598--606, 2016.

\bibitem[IO19]{ito2019reconfiguration}
Takehiro Ito and Yota Otachi.
\newblock Reconfiguration of colorable sets in classes of perfect graphs.
\newblock {\em Theor. Comput. Sci.}, 772:111--122, 2019.

\bibitem[JG79]{johnson1979computers}
David~S. Johnson and Michael~R. Garey.
\newblock {\em Computers and Intractability: A Guide to the Theory of
  {NP}-Completeness}.
\newblock W. H. Freeman, 1979.

\bibitem[KKT03]{kempe2003maximizing}
David Kempe, Jon Kleinberg, and {\'{E}}va Tardos.
\newblock Maximizing the spread of influence through a social network.
\newblock In {\em KDD}, pages 137--146, 2003.

\bibitem[KLV17]{kyncl2017irreversible}
Jan Kyn{\v{c}}l, Bernard Lidick{\'y}, and Tom{\'a}{\v{s}} Vysko{\v{c}}il.
\newblock Irreversible 2-conversion set in graphs of bounded degree.
\newblock {\em Discrete Math. Theor. Comput. Sci.}, 19(3), 2017.

\bibitem[KMM11]{kaminski2011shortest}
Marcin Kami{\'n}ski, Paul Medvedev, and Martin Milani{\v{c}}.
\newblock Shortest paths between shortest paths.
\newblock {\em Theor. Comput. Sci.}, 412(39):5205--5210, 2011.

\bibitem[KMM12]{kaminski2012complexity}
Marcin Kami{\'n}ski, Paul Medvedev, and Martin Milani{\v{c}}.
\newblock Complexity of independent set reconfigurability problems.
\newblock {\em Theor. Comput. Sci.}, 439:9--15, 2012.

\bibitem[LM19]{lokshtanov2019complexity}
Daniel Lokshtanov and Amer~E. Mouawad.
\newblock The complexity of independent set reconfiguration on bipartite
  graphs.
\newblock {\em ACM Trans. Algorithms}, 15(1):7:1--7:19, 2019.

\bibitem[MNPR17]{mouawad2017shortest}
Amer~E. Mouawad, Naomi Nishimura, Vinayak Pathak, and Venkatesh Raman.
\newblock Shortest reconfiguration paths in the solution space of {Boolean}
  formulas.
\newblock {\em {SIAM} J. Discret. Math.}, 31(3):2185--2200, 2017.

\bibitem[MNR{\etalchar{+}}17]{mouawad2017parameterized}
Amer~E. Mouawad, Naomi Nishimura, Venkatesh Raman, Narges Simjour, and Akira
  Suzuki.
\newblock On the parameterized complexity of reconfiguration problems.
\newblock {\em Algorithmica}, 78(1):274--297, 2017.

\bibitem[MNRS18]{mouawad2018vertex}
Amer~E. Mouawad, Naomi Nishimura, Venkatesh Raman, and Sebastian Siebertz.
\newblock Vertex cover reconfiguration and beyond.
\newblock {\em Algorithms}, 11(2):20, 2018.

\bibitem[MNRW14]{mouawad2014reconfiguration}
Amer~E. Mouawad, Naomi Nishimura, Venkatesh Raman, and Marcin Wrochna.
\newblock Reconfiguration over tree decompositions.
\newblock In {\em IPEC}, pages 246--257, 2014.

\bibitem[Moh01]{mohar2001face}
Bojan Mohar.
\newblock Face covers and the genus problem for apex graphs.
\newblock {\em J. Comb. Theory, Ser. {B}}, 82(1):102--117, 2001.

\bibitem[Mou15]{mouawad2015reconfiguration}
Amer Mouawad.
\newblock {\em On Reconfiguration Problems: Structure and Tractability}.
\newblock PhD thesis, University of Waterloo, 2015.

\bibitem[MTY11]{makino2011exact}
Kazuhisa Makino, Suguru Tamaki, and Masaki Yamamoto.
\newblock An exact algorithm for the {Boolean} connectivity problem for
  $k$-{CNF}.
\newblock {\em Theor. Comput. Sci.}, 412(35):4613--4618, 2011.

\bibitem[Mun17]{munaro2017line}
Andrea Munaro.
\newblock On line graphs of subcubic triangle-free graphs.
\newblock {\em Discrete Math.}, 340(6):1210--1226, 2017.

\bibitem[Nis18]{nishimura2018introduction}
Naomi Nishimura.
\newblock Introduction to reconfiguration.
\newblock {\em Algorithms}, 11(4):52, 2018.

\bibitem[NNUW13]{nichterlein2013tractable}
Andr{\'e} Nichterlein, Rolf Niedermeier, Johannes Uhlmann, and Mathias Weller.
\newblock On tractable cases of target set selection.
\newblock {\em Social Netw. Analys. Mining}, 3(2):233--256, 2013.

\bibitem[Pel98]{peleg1998size}
David Peleg.
\newblock Size bounds for dynamic monopolies.
\newblock {\em Discrete Appl. Math.}, 86(2-3):263--273, 1998.

\bibitem[Pel02]{peleg2002local}
David Peleg.
\newblock Local majorities, coalitions and monopolies in graphs: A review.
\newblock {\em Theory Comput. Syst.}, 282(2):231--257, 2002.

\bibitem[PPRS14]{penso2014p3}
Lucia~Draque Penso, F{\'a}bio Protti, Dieter Rautenbach, and U{\'e}verton~S.
  Souza.
\newblock On {$P_3$}-convexity of graphs with bounded degree.
\newblock In {\em AAIM}, pages 263--274, 2014.

\bibitem[TU15]{takaoka2015note}
Asahi Takaoka and Shuichi Ueno.
\newblock A note on irreversible 2-conversion sets in subcubic graphs.
\newblock {\em {IEICE} Trans. Inf. Syst.}, 98(8):1589--1591, 2015.

\bibitem[UKG88]{ueno1988nonseparating}
Shuichi Ueno, Yoji Kajitani, and Shin'ya Gotoh.
\newblock On the nonseparating independent set problem and feedback set problem
  for graphs with no vertex degree exceeding three.
\newblock {\em Discrete Math.}, 72(1-3):355--360, 1988.

\bibitem[vdH13]{heuvel13complexity}
Jan van~den Heuvel.
\newblock The complexity of change.
\newblock In {\em Surveys in Combinatorics 2013}, volume 409, pages 127--160.
  Cambridge University Press, 2013.

\bibitem[Wro18]{wrochn2018reconfiguration}
Marcin Wrochna.
\newblock Reconfiguration in bounded bandwidth and treedepth.
\newblock {\em J. Comput. Syst. Sci.}, 93:1--10, 2018.

\end{thebibliography}

\appendix
\section{Missing Proofs}
\label{app:proofs}

\begin{proof}[Proof of \cref{thm:33}]

We reduce from \prb{Minimum Vertex Cover Reconfiguration} on graphs of degree $2$ and $3$.
To this end, we introduce the following gadget (see \cref{fig:sigma}):
\begin{shadebox}
\centering
\textbf{Construction of $\Sigma$-gadget (\cref{fig:sigma}).}
\begin{description}
\item[\textbf{Step 1.}] Create a vertex set $V \triangleq \{r,t_1,t_2,t_3,t_4\}$.
\item[\textbf{Step 2.}] Create an edge set
$E \triangleq \{(r,t_1), (r,t_3), (t_1,t_2), (t_1,t_4), (t_2, t_3), (t_2, t_4), (t_3,t_4)\}$.
\end{description}
\end{shadebox}
\noindent
Similar gadgets can be found in \cite{hearn2005pspace,hearn2009games}.
We here call this gadget a \emph{$\Sigma$-gadget}.
Observe that a $\Sigma$-gadget is planar.
We say that a $\Sigma$-gadget is \emph{connected to} vertex $v$
if there exists an edge $(r,v)$.
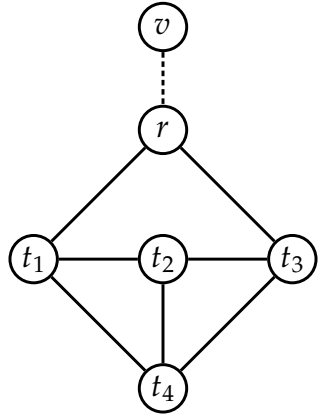
\begin{figure}[tbp]
\centering
\scalebox{0.7}{
    \begin{tikzpicture}
    [circlenode/.style={draw, circle, minimum height=1cm, font=\LARGE}]
    \node[mynode](v){$v$};
    \node[mynode, below=1cm of v](r){$r$};
    \node[mynode, below=1.5cm of r](t2){$t_2$};
    \node[mynode, left=1.5cm of t2](t1){$t_1$};
    \node[mynode, right=1.5cm of t2](t3){$t_3$};
    \node[mynode, below=1.5cm of t2](t4){$t_4$};
    
    \foreach \u / \v in {r/t1, r/t3, t1/t2, t2/t3, t1/t4, t2/t4, t3/t4}
        \draw[myedge] (\u)--(\v);
    \draw[myedge, densely dashed] (v)--(r);
    \end{tikzpicture}
}
\caption{$\Sigma$-gadget.}
\label{fig:sigma}
\end{figure}

\cref{fig:sigma-vc} lists minimum vertex covers of a $\Sigma$-gadget.
We define $M \triangleq \{r,t_2,t_4\}$, which is the minimum vertex cover drawn in \cref{fig:sigma-vc1}.
We then have the following.

\begin{lemma}
\label{lem:sigma}
Let $G=(V,E)$ be a graph and $G' = (V',E')$ be a graph
obtained from $G$ by connecting a $\Sigma$-gadget $R$ to a vertex $v$ of $G$.
Then,
a vertex set $S \subseteq V$ is a minimum vertex cover of $G$
if and only if
$S\uplus M$ is a minimum vertex cover of $G'$.
Moreover, two minimum vertex covers $X$ and $Y$ of $G$ are \TJ-reconfigurable on $G$
if and only if $X \uplus M$ and $Y\uplus M$ are \TJ-reconfigurable on $G'$.
\end{lemma}
\begin{proof}
Observing the following facts suffices to ensure the statement:
\begin{description}
\item[\textbf{(1)}]
if $S \subseteq V$ is a minimum vertex cover of $G$,
then $S \uplus M$ is a minimum vertex cover of $G'$;
\item[\textbf{(2)}]
if $S' \subseteq V'$ is a minimum vertex cover of $G'$,
then $S' \cap V$ is a minimum vertex cover of $G$ and
$S' \cap V(R)$ is a minimum vertex cover of $R$;
\item[\textbf{(3)}]
$M$ is not \TJ-reconfigurable to any other minimum vertex cover of $R$ (see \cref{fig:sigma-vc}).
\end{description}
\end{proof}

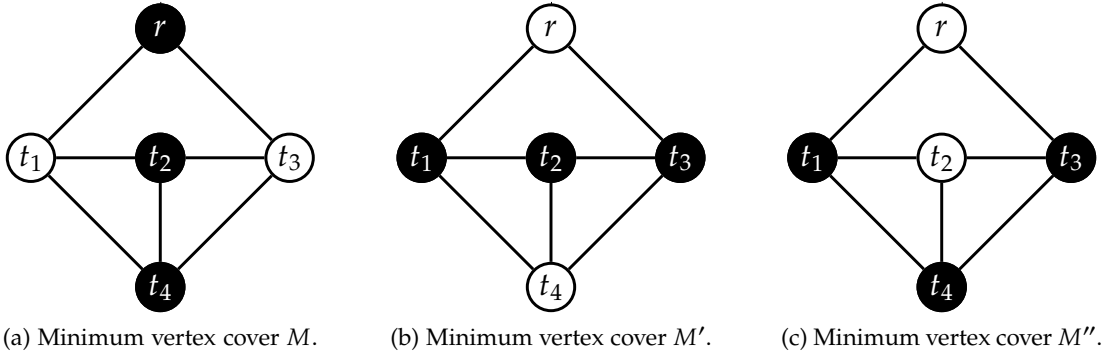
\begin{figure}[tbp]
    \centering
    \null\hfill
    \subfloat[Minimum vertex cover $M$.\label{fig:sigma-vc1}]{\scalebox{0.7}{
    \begin{tikzpicture}
        \node[mynode](r){$r$};
        \node[mynode, below=1.5cm of r](t2){$t_2$};
        \node[mynode, left=1.5cm of t2](t1){$t_1$};
        \node[mynode, right=1.5cm of t2](t3){$t_3$};
        \node[mynode, below=1.5cm of t2](t4){$t_4$};

        \node[mytoken]() at (r){$r$};
        \node[mytoken]() at (t2){$t_2$};
        \node[mytoken]() at (t4){$t_4$};
        
        \foreach \u / \v in {r/t1, r/t3, t1/t2, t2/t3, t1/t4, t2/t4, t3/t4}
            \draw[myedge] (\u)--(\v);
        \draw[myedge, densely dashed] (v)--(r);
    \end{tikzpicture}
    }}
    \hfill
    \subfloat[Minimum vertex cover $M'$.]{\scalebox{0.7}{
    \begin{tikzpicture}
        \node[mynode](r){$r$};
        \node[mynode, below=1.5cm of r](t2){$t_2$};
        \node[mynode, left=1.5cm of t2](t1){$t_1$};
        \node[mynode, right=1.5cm of t2](t3){$t_3$};
        \node[mynode, below=1.5cm of t2](t4){$t_4$};
        
        \node[mytoken]() at (t1){$t_1$};
        \node[mytoken]() at (t2){$t_2$};
        \node[mytoken]() at (t3){$t_3$};

        \foreach \u / \v in {r/t1, r/t3, t1/t2, t2/t3, t1/t4, t2/t4, t3/t4}
            \draw[myedge] (\u)--(\v);
        \draw[myedge, dashed] (v)--(r);
    \end{tikzpicture}
    }}
    \hfill
    \subfloat[Minimum vertex cover $M''$.]{\scalebox{0.7}{
    \begin{tikzpicture}
        \node[mynode](r){$r$};
        \node[mynode, below=1.5cm of r](t2){$t_2$};
        \node[mynode, left=1.5cm of t2](t1){$t_1$};
        \node[mynode, right=1.5cm of t2](t3){$t_3$};
        \node[mynode, below=1.5cm of t2](t4){$t_4$};
        
        \node[mytoken]() at (t1){$t_1$};
        \node[mytoken]() at (t4){$t_4$};
        \node[mytoken]() at (t3){$t_3$};

        \foreach \u / \v in {r/t1, r/t3, t1/t2, t2/t3, t1/t4, t2/t4, t3/t4}
            \draw[myedge] (\u)--(\v);
        \draw[myedge, dashed] (v)--(r);
    \end{tikzpicture}
    }}
    \hfill\null
    \caption{Three minimum vertex covers of a $\Sigma$-gadget, denoted by black circles $\bullet$.}
    \label{fig:sigma-vc}
\end{figure}

The reduction from a planar graph of degree $2$ and $3$ to a planar $3$-regular graph is presented below.
\begin{shadebox}
\centering
\textbf{Reduction from planar graph $G$ of degree $2$ and $3$ to planar $3$-regular graph $H$.}
\begin{description}
\item[\textbf{Step 1.}] Connect a $\Sigma$-gadget to each degree-2 vertex $v$ of $G$ to obtain a graph $H$.
Let $M_v$ denote the minimum vertex cover of the $\Sigma$-gadget connected to vertex $v$ defined above.
\end{description}
\end{shadebox}
\noindent
Obviously, the reduction completes in polynomial time, and
$H$ is a planar $3$-regular graph.
Let $X$ and $Y$ be two minimum vertex covers of $G$.
By applying \cref{lem:sigma} repeatedly, we have that
$X$ and $Y$ are \TJ-reconfigurable on $G$ if and only if
$X \uplus \bigcup_{v} M_v$ and $Y \uplus \bigcup_{v} M_v$ are \TJ-reconfigurable on $H$.
Since \prb{Minimum Vertex Cover Reconfiguration} on planar graphs of degree $2$ and $3$ is \PSPACE-complete \cite{hearn2005pspace},
we obtain the desired result.
\end{proof}

\begin{proof}[Proof of \cref{lem:mts-theta}]
One can verify that the seed set $\{r,t_{1,2},t_{2,3}\}$ is a target set of $R$,
which is drawn in \cref{fig:theta-mts}:
$r$, $t_{1,2}$, and $t_{2,3}$ become activated initially;
$t_{1,1}$, $t_{1,3}$, and $t_{2,2}$ become activated at step $1$;
$t_{2,1}$ becomes activated at step $2$;
$t_{2,6}$ becomes activated at step $3$;
$t_{1,6}$ becomes activated at step $4$;
$t_{1,5}$ becomes activated at step $5$;
$t_{2,5}$ and $t_{1,4}$ become activated at step $6$;
$t_{2,4}$ becomes activated at step $7$.
Showing that any size-$2$ seed set is not a target set of $R'$ is sufficient to prove the statement.
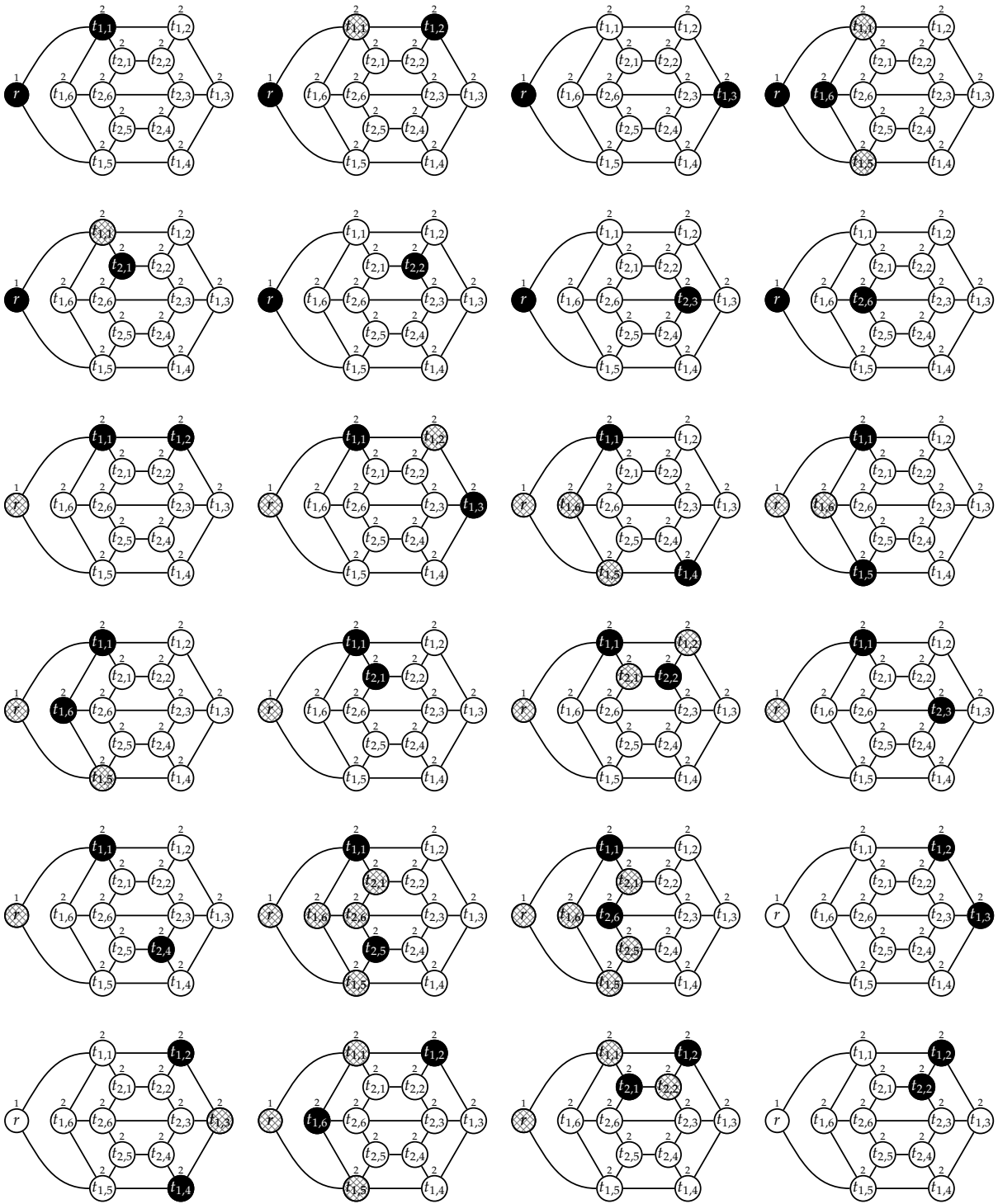
To this end, we exhaustively enumerate all possible size-$2$ seed sets in \cref{fig:theta-lemma-1,fig:theta-lemma-2},
where a black circle $\bullet$ denotes a seed, and
a cross-hatched circle denotes an activated vertex.
Note that we have omitted some seed sets that are identical
due to the symmetry of $R'$; e.g.,
$\{t_{1,1}, t_{2,6}\}$ and $\{t_{1,5}, t_{2,6}\}$ are identical.
\end{proof}
\input{fig_theta-lemma}

\end{document}